\documentclass[prd,amsfonts,aps,nofootinbib,notitlepage,11pt,superscriptaddress]{revtex4-1}
\pdfoutput=1
\usepackage{amsmath,amssymb}
\usepackage{graphicx}
\usepackage[utf8]{inputenc}
\usepackage{cancel}
\usepackage[colorlinks=true]{hyperref}
\usepackage{color}
\usepackage{array,multirow}
\usepackage{subcaption}
\usepackage{float}
\usepackage{soul}

\newcommand{\UPTC}{Escuela de Física, Universidad Pedagógica y Tecnológica de Colombia,\\
Avenida Central del Norte \# 39-115, Tunja, Colombia}
\newcommand{\UdeA}{Instituto de Física, Universidad de Antioquia,\\Calle 70 \# 52-21, Apartado Aéreo 1226, Medellín, Colombia}
\newcommand{\Lapth}{LAPTh, CNRS, USMB, 9 Chemin de Bellevue, 74940 Annecy, France}
\newcommand{\skobe}{Skobeltsyn Institute of Nuclear Physics, Moscow State University, Moscow 119992, Russia}

\begin{document}
\title{The  $Z_7$ model of three-component scalar dark matter}

\author{Genevi\`eve B\'elanger}\affiliation{\Lapth}
\author{Alexander Pukhov} \affiliation{\skobe}
\author{Carlos E. Yaguna}
\affiliation{\UPTC}
\author{\'Oscar Zapata}
\affiliation{\UdeA}
          
\begin{abstract}
We investigate, for the first time, a scenario where the dark matter consists of three complex scalar fields that are stabilized by a single $Z_7$ symmetry. As an extension of the well-known scalar Higgs-portal, this $Z_7$ model is also subject to important restrictions arising from the relic density constraint and from direct detection experiments. Our goal in this paper is to find and characterize the  viable regions of this model, and to analyze its detection prospects in future experiments. First, the processes that affect the relic densities are identified (they include semiannihilations and conversions) and then incorporated into  the Boltzmann equations for the dark matter abundances, which are numerically 
solved with micrOMEGAs. By means of random scans of the parameter space,  the regions consistent with current data, including the recent direct detection limit from the LZ experiment, are selected.  Our results reveal that the $Z_7$ model is indeed viable over a wide range of dark matter masses and that both conversions and semiannihilations play an important role in determining the relic densities. Remarkably,  we find that in many cases all three of the dark matter particles give rise to observable signals in future direct detection experiments, providing a suitable way to test this scenario. 
\end{abstract}

\maketitle
\section{Introduction}
Even though dark matter accounts for about $25\%$ of the energy-density of the Universe \cite{Planck:2018vyg}, its fundamental nature remains unknown and constitutes one of the most important open problems in astro-particle physics today.  The most common hypothesis is that it consists of a \emph{single} neutral and stable elementary particle not present in the Standard Model~\cite{Jungman:1995df,Bertone:2004pz}. Typical WIMPs~\cite{Roszkowski:2017nbc,Arcadi:2017kky}, such as the neutralino~\cite{Ellis:1983ew} or the singlet scalar~\cite{Silveira:1985rk,McDonald:1993ex,Burgess:2000yq}, are explicit realizations of this idea. Another possibility compatible with current data is that the dark matter consists of \emph{several} WIMPs, each contributing a fraction of the observed dark matter density~\cite{Boehm:2003ha,Ma:2006uv,Cao:2007fy,Hur:2007ur,Lee:2008pc,Zurek:2008qg,Barger:2008jx,Profumo:2009tb,Batell:2010bp,Belanger:2011ww,Liu:2011aa,Ivanov:2012hc,Belanger:2012vp,Modak:2013jya,Belanger:2014vza,Esch:2014jpa,Belanger:2014bga,Cai:2015zza,Biswas:2015sva,Arcadi:2016kmk,Bhattacharya:2016ysw,Bhattacharya:2017fid,Ahmed:2017dbb,Bhattacharya:2018cgx,YaserAyazi:2018lrv,Bernal:2018aon,Poulin:2018kap,Borah:2019aeq,Nanda:2019nqy,Yaguna:2019cvp,Betancur:2020fdl,Hernandez-Sanchez:2020aop,Belanger:2020hyh,Choi:2021yps,Belanger:2021lwd,Yaguna:2021vhb,DiazSaez:2021pmg,Mohamadnejad:2021tke,Chakrabarty:2021kmr,Ho:2022erb,Bhattacharya:2022wtr,DuttaBanik:2020jrj,Hernandez-Sanchez:2022dnn,BasiBeneito:2020mdr,Hall:2019rld,Hall:2021zsk,Das:2022oyx}. Compelling realizations of these \emph{multi-component} dark matter scenarios can be obtained when the usual $Z_2$ stabilizing symmetry is replaced by a $Z_N$, with  $N\geq 3$~\cite{Batell:2010bp,Belanger:2012vp,Belanger:2014bga,Yaguna:2019cvp}. 

Recently, detailed phenomenological analyses of two-component dark matter models based on $Z_4$, $Z_5$, and $Z_6$ symmetries were reported \cite{Belanger:2020hyh,Yaguna:2021vhb,Yaguna:2021rds}. Depending on the symmetry, the dark matter particles could be two complex scalars ($Z_5$), a complex scalar and a real scalar ($Z_4$ or $Z_6$), or a  real scalar and a fermion ($Z_4$). All these models can be seen as natural extensions of the well-known scalar Higgs-portal~\cite{Patt:2006fw}, but they have the advantage of remaining viable for dark matter masses below $1-2$ TeV.  Indeed in  these models new interactions allowed by $Z_N$ symmetries strongly impact DM formation, namely  semi-annihilation processes involving the annihiliation of a pair of DM particles into a single DM and a SM particle~\cite{Hambye:2008bq,Hambye:2009fg,DEramo:2010keq,Belanger:2012vp} as well as conversion processes involving different DM particles ~\cite{Belanger:2011ww,Arcadi:2016kmk,Belanger:2020hyh}. Moreover these models  provide distinctive experimental signatures that may enable to differentiate them from the more conventional scenarios with just one dark matter particle. So far, though, the feasibility and testability of $Z_N$ scenarios with more than two dark matter particles remains an open question. 

Three-component dark matter scenarios are qualitatively different from two-component ones due to the presence of novel processes affecting simultaneously the relic densities of the three dark matter particles. Among them, let us mention semiannihilation processes of the type $\phi_i+\phi_j\to \phi_k^\dagger +h$ and conversion processes such as $\phi_i+\phi_i \to \phi_j^\dagger+\phi_k^\dagger$, with $i\neq j \neq k$. As we will show, these processes, which have no analogue within two-component dark matter frameworks and had not been taken into account before, are quantitatively important for the determination of the relic densities and, consequently, of the viable regions of $Z_N$ scenarios.  As a case in point, the $Z_7$ model for three-component scalar dark matter is extremely appealing, not only because  a $Z_7$  is the lowest $Z_N$ symmetry that can stabilize three complex scalar fields, but also because such model serves as a prototype for other scenarios with three dark matter particles but a different $Z_N$ symmetry.

In this paper, we analyze, for the first time, the three-component dark matter model based on the $Z_7$ symmetry. In it, the dark matter is explained by  three complex scalar fields that are SM singlets but  transform non-trivially (and differently) under the $Z_7$. Aside from  the dark matter particles, the model does not include any additional fields. Our aim is to characterize the  regions of this model that are consistent with current data, and to investigate its detectability in future experiments.  To that end, a complete calculation of the relic densities, including all the interactions between the different DM components in the early Universe, will be presented. We will see that this $Z_7$ model is viable over a wide range of dark  matter masses, and that its direct detection prospects are excellent. Moreover, many of our results can be generalized to scenarios with other $Z_N$ symmetries.

In the literature it is possible to find different scenarios with three dark matter particles, but most of them rely on three independent $Z_2$ symmetries to stabilize them.  Among them, let us mention the following alternatives for the dark matter particles: one vector and 2 fermions \cite{Ahmed:2017dbb}, one vector or a scalar and at least two scalars \cite{Poulin:2018kap,ChuliaCentelles:2022ogm}, one vector, one fermion and one scalar \cite{Elahi:2019jeo}, two fermions and one scalar \cite{Bhattacharya:2018cgx}, and three scalars \cite{Choi:2021yps}. The main novelty of our model is thus the much simpler assumption of a single $Z_7$ symmetry, which  may be a remnant of a $U(1)$ gauge symmetry that has been spontaneously broken at higher energies.  Such a symmetry entails novel conversion and semiannihilation processes among the dark matter particles that affect their relic densities,  opening up new viable regions of parameter space. Finally, let us mention that this study provides the first detailed application of a new version of micrOMEGAs for N-component dark matter that will be released shortly.

The rest of the paper is organized as follows:  the model is introduced in the next section while its dark matter phenomenology is described in section \ref{sec:pheno}; our main results are presented in section \ref{sec:viable}, where the results of different scans of the parameter space are analyzed; in section \ref{sec:disc} our main findings are contrasted  against those from other $Z_N$ models; finally our conclusions are drawn in section \ref{sec:conc}.

\section{The $Z_7$ model}
\label{sec:model}

\begin{figure}
\centering
\includegraphics[scale=0.6]{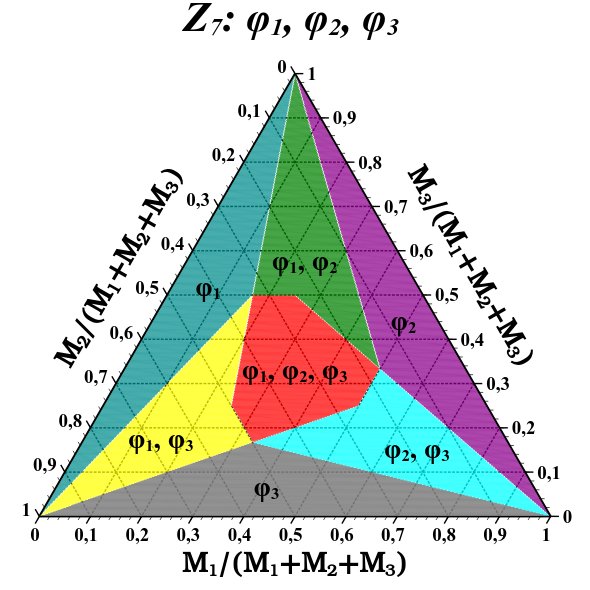}\\
\caption{A ternary plot of the stability regions for the $Z_7$ model. For each region, the fields that are stable within it are explicitly written. The three-component dark matter scenario that is of interest to us is realized in the central (red) region.}
\label{fig:trianglep123}
\end{figure}

Let us consider a scenario with three new complex scalar fields, $\phi_{1,2,3}$, charged under a  $Z_7$ symmetry. The unique charge assignment (up to trivial field redefinitions) that allows all three fields to be stable is \cite{Yaguna:2019cvp} 
\begin{align}
    \phi_1\sim \omega_7,\,\,\, \phi_2\sim \omega_7^2,\,\,\, \phi_3\sim \omega_7^3; \hspace{1cm}\omega_7=\exp(i2\pi/7). 
\end{align}
These new fields--the dark matter particles--are assumed to be  singlets of the SM gauge group whereas the SM particles are taken to be singlets under the $Z_7$.  The most general $Z_7$-invariant scalar potential is then given by 
\begin{align}\label{eq:Z7lag}
 \mathcal{V}&=\,\,\mu^2_H|H|^2+\lambda_H|H|^4+\sum_{i=1}^3\left[\mu_{i}^2|\phi_i|^2+\lambda_{4i}|\phi_i|^4+\lambda_{Si}|H|^2|\phi_i|^2\right]\nonumber\\
 & \,+\lambda_{412}|\phi_1|^2|\phi_2|^2+\lambda_{413}|\phi_1|^2|\phi_3|^2+\lambda_{423}|\phi_2|^2|\phi_3|^2\nonumber\\
 &\,+\frac{1}{2}\left[\mu_{S1}\phi^2_1\phi_2^{*} + \mu_{S2}\phi_2^2\phi_3 + \mu_{S3}\phi_3^2\phi_1 + 2\mu_{S4}\phi_1\phi_2\phi_3^* + \text{H.c.}\right]\nonumber\\
 &\,+\left[\lambda_{31}\phi _1 \phi _2^3+\lambda_{32}\phi _1^2 \phi _2\phi_3+\lambda_{33}\phi _1^* \phi _2\phi_3^2 + \lambda_{34}\phi _1^{*3} \phi_3 +\lambda_{35} \phi_2^{*}\phi_3^3 +\lambda_{36}\phi _1 \phi _2^{*2}\phi_3 + \text{H.c.}\right],
 \end{align}
where $H$ is the SM Higgs doublet.  Notice that  this model can  describe scenarios with one, two, or three dark matter particles, depending on the relations among the  masses of the new scalar fields. Figure \ref{fig:trianglep123} displays, in a ternary plot, the regions consistent with these different possibilities. The case of a single dark matter particle is rather similar to the scalar complex singlet model~\cite{Silveira:1985rk,McDonald:1993ex,Burgess:2000yq} whereas the two-component scenario shares most of the features of the $Z_5$ model~\cite{Belanger:2020hyh}. In this work, we are interested only in the case where  a three-component dark matter scenario is realized --the central red region in figure \ref{fig:trianglep123}. Therefore, we assume in the following that $\phi_{1,2,3}$ do not acquire a vacuum expectation value and  that their masses  satisfy $M_2<2 M_1,\, M_3<2 M_2,\, M_1<2 M_3,\, M_1<M_2+M_3,\, M_2<M_1+M_3,\,M_3<M_1+M_2$. These conditions are obtained directly from equation (\ref{eq:Z7lag}) and guarantee the stability of $\phi_{1}$, $\phi_2$ and $\phi_3$.

 \begin{figure}
\centering
\includegraphics[scale=1]{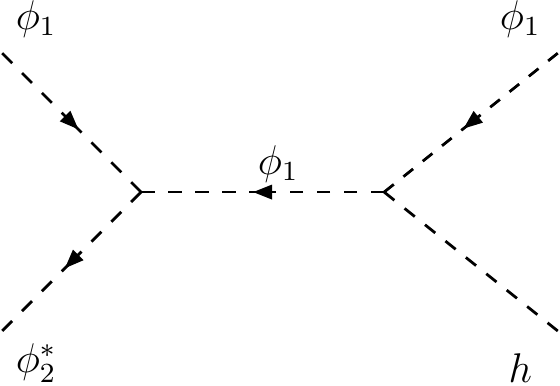}\\
\includegraphics[scale=1]{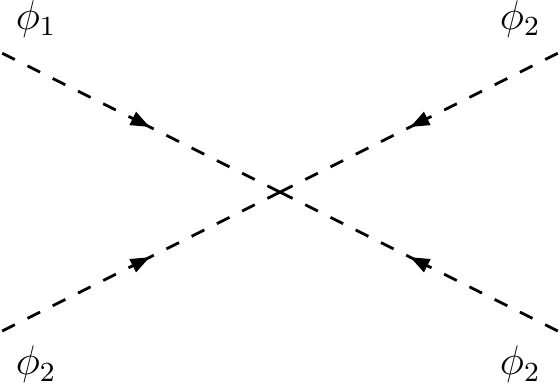}\hspace{1cm}
\includegraphics[scale=1]{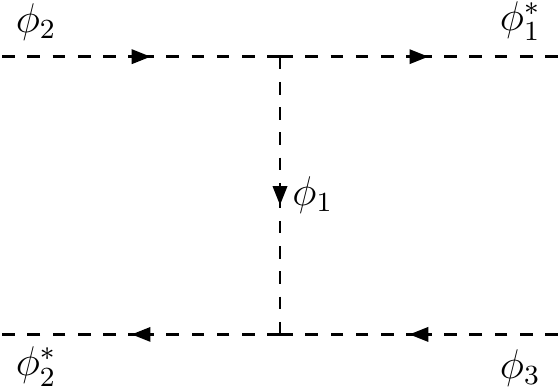}
\caption{Some examples of semiannihilation (top) and conversion (bottom) processes allowed in the $Z_7$ model.}
\label{fig:processes}
\end{figure}

In total, this model contains $22$ new parameters ($7$ dimensionful and $15$ dimensionless), but three of them --$\lambda_{4i}$-- are irrelevant for the dark matter phenomenology and can be ignored in our analysis. The parameters $\mu_i^2$ ($i=1,2,3$), on the other hand, can be conveniently traded for the physical masses $M_i$ of the scalar fields,  so that the free parameters of the model may be taken to be $M_i$ (3), $\lambda_{Si}$ (3),  $\lambda_{4ij}$ (3), $\mu_{Si}$ (4), and $\lambda_{3i}$ (6).  Without loss of  generality, we will  restrict our numerical analysis to the hierarchy $M_1<M_2<M_3$ while leaving the couplings free.  For simplicity,  all couplings are  taken to be real.

Our goal in this paper is to study the dark matter phenomenology of this model. In particular, we will identify and characterize the regions, within its multi-dimensional parameter space, that lead to scenarios consistent with all present data, including the  relic density and the limits from dark matter experiments. In addition, we will show that current and future direct detection experiments offer a promising way to test this scenario.

\section{Dark Matter Phenomenology}
\label{sec:pheno}
In this model, the dark matter particles, $\phi_{1,2,3}$, couple to the SM fields only via the Higgs-portal --see the $\lambda_{Si}$ terms in \ref{eq:Z7lag}. In addition, they couple among themselves through the different trilinear and quartic terms allowed by the $Z_7$ symmetry. As a consequence, a large number of $2\to 2$ processes can contribute to their relic densities --see figure \ref{fig:processes} for some examples. Table \ref{tab:processes} shows the complete list of processes for each of the  dark matter particles, classified according to their type. They include, besides the usual annihilations into SM particles, several examples of semiannihilations and of dark matter conversions, which are typical of $Z_N$ models \cite{Yaguna:2019cvp}.  Half of the different types of processes, in fact,  had not been studied before, as they involve three different dark matter particles. Notice that semiannihilations always involve a combination of a trilinear interaction with the  higgs-portal (top diagram in figure \ref{fig:processes}), whereas dark matter conversions can be obtained either via a single quartic interaction or  two trilinear ones (bottom diagrams in figure \ref{fig:processes}). 
\begin{table}[t]
    
    \begin{tabular}{c |c}
      $\phi_1$ Processes   & Type  \\
      \hline 
        $\phi_1+\phi_1^\dagger\to SM + SM$ & $1100$\\
        $\phi_1+\phi_1^\dagger \to \phi_2+\phi_2^\dagger$  & $1122$\\
        $\phi_1+\phi_1^\dagger \to \phi_3+\phi_3^\dagger$  & $1133$\\
        $\phi_1+\phi_1 \to \phi_2+h$  & $1120$\\
        $\phi_1+\phi_1 \to \phi_2^\dagger+\phi_3^\dagger$  & $1123$\\
        $\phi_1+\phi_1 \to \phi_1^\dagger+\phi_3$  & $1113$ \\
        $\phi_1+h \to \phi_3^\dagger +\phi_3^\dagger $  & $1033$\\
        $\phi_1+h \to \phi_3 +\phi_2^\dagger $  & $1023$\\
        $\phi_1 + \phi_2 \to \phi_3+h $ & $1230$\\
        $\phi_1 + \phi_2  \to \phi_2^\dagger+\phi_2^\dagger $ & $1222$\\
        $\phi_1 + \phi_2^\dagger  \to \phi_3+\phi_3 $ & $1233$\\
        $\phi_1 + \phi_2^\dagger  \to \phi_3^\dagger+\phi_2 $ & $1223$\\
        $\phi_1+\phi_3 \to \phi_2 + \phi_2$ & $1322$\\
        $\phi_1+\phi_3 \to \phi_3^\dagger + h$ & $1330$\\
        $\phi_1+\phi_3^\dagger \to \phi_2^\dagger + h$ & $1320$\\
        $\phi_1+\phi_3^\dagger \to \phi_2 + \phi_3$ & $1323$\\
    \end{tabular}\hspace{5mm}
    \begin{tabular}{c |c}
      $\phi_2$ Processes   & Type  \\
      \hline 
        $\phi_2+\phi_2^\dagger\to SM + SM$ & $2200$\\
        $\phi_2+\phi_2^\dagger \to \phi_1+\phi_1^\dagger$  & $2211$\\
        $\phi_2+\phi_2^\dagger \to \phi_3+\phi_3^\dagger$  & $2233$\\
        $\phi_2+\phi_2 \to \phi_3^\dagger+h$  & $2230$\\
        $\phi_2+\phi_2 \to \phi_1+\phi_3$  & $2213$\\
        $\phi_2+\phi_2 \to \phi_1^\dagger+\phi_2^\dagger$  & $2212$\\
        $\phi_2+h \to \phi_1 +\phi_1 $  & $2011$\\
        $\phi_2+h \to \phi_1^\dagger +\phi_3 $  & $2013$\\
        $\phi_2 + \phi_1 \to \phi_3+h $ & $1230$\\
        $\phi_2 + \phi_1  \to \phi_3^\dagger+\phi_1^\dagger $ & $1213$\\
        $\phi_2 + \phi_1^\dagger  \to \phi_1+h $ & $1210$\\
        $\phi_2 + \phi_1^\dagger  \to \phi_3^\dagger+\phi_3^\dagger $ & $1233$\\
        $\phi_2+\phi_3 \to \phi_1^\dagger + \phi_1^\dagger$ & $2311$\\
        $\phi_2+\phi_3 \to \phi_3^\dagger + \phi_1$ & $2313$\\
        $\phi_2+\phi_3^\dagger \to \phi_1^\dagger + h$ & $2310$\\
        $\phi_2+\phi_3^\dagger \to \phi_3 + \phi_3$ & $2333$\\
    \end{tabular}\hspace{5mm}
         \begin{tabular}{c |c}
      $\phi_3$ Processes   & Type  \\
      \hline 
        $\phi_3+\phi_3^\dagger\to SM + SM$ & $3300$\\
        $\phi_3+\phi_3^\dagger \to \phi_1+\phi_1^\dagger$  & $3311$\\
        $\phi_3+\phi_3^\dagger \to \phi_2+\phi_2^\dagger$  & $3322$\\
        $\phi_3+\phi_3 \to \phi_1^\dagger+h$  & $3310$\\
        $\phi_3+\phi_3 \to \phi_2^\dagger+\phi_1$  & $3312$\\
        $\phi_3+\phi_3 \to \phi_3^\dagger+\phi_2$  & $3323$\\
        $\phi_3+h \to \phi_1 +\phi_2 $  & $3012$\\
        $\phi_3+h \to \phi_2^\dagger +\phi_2^\dagger $  & $3022$\\
        $\phi_3 + \phi_1 \to \phi_2+\phi_2 $ & $1322$\\
        $\phi_3 + \phi_1  \to \phi_2^\dagger+\phi_1^\dagger $ & $1312$\\
        $\phi_3 + \phi_1^\dagger  \to \phi_2+h $ & $1320$\\
        $\phi_3 + \phi_1^\dagger  \to \phi_1+\phi_1 $ & $1311$\\
        $\phi_3+\phi_2 \to \phi_1^\dagger + \phi_1^\dagger$ & $2311$\\
        $\phi_3+\phi_2 \to \phi_2^\dagger + h$ & $2320$\\
        $\phi_3+\phi_2^\dagger \to \phi_1 + h$ & $2310$\\
        $\phi_3+\phi_2^\dagger \to \phi_1 + \phi_2^\dagger$ & $2312$\\
    \end{tabular}
    
    \caption{The $2\to 2$ processes that are allowed in the $Z_7$ model and that can modify the relic density of $\phi_1$ (left), $\phi_2$ (center) or $\phi_3$ (right). $h$ denotes the SM Higgs boson. Notice that half of the processes types cannot be found within two-componet dark matter scenarios, as they involve three different dark matter particles. Conjugate and inverse processes are not shown.  
    }
    \label{tab:processes}
\end{table}

All these $2\to 2$ processes must be taken into account in the set of Boltzmann equations that determine the evolution of the dark matter abundances and, ultimately, the  relic densities. Let us now present the equations for the evolution of  DM abundances in generic multi-component scenarios, written in a way that allows a simple numerical implementation.  

We assume that all DM particles are split into  sectors,   labelled  by letters of the Greek   alphabet,  and that chemical equilibrium is maintained within each sector. The $Z_7$ model is particularly simple because it consists of three dark matter sectors, each containing just one dark matter particle.  In this paper we assume kinetic equilibrium of all DM particles with SM bath particles,  so  the momentum distribution of all  DM particles is described by the bath  temperature $T$, but the number density can deviate from the thermal equilibrium case and this deviation is described  by a common factor in each sector.

As usual,  we write  equations  for the abundances 
\begin{equation}
 Y_{\alpha}=n_{\alpha}/\mathfrak{s}(T)
\end{equation} 
where $n_{\alpha}$ is a particle number density and $\mathfrak{s}(T)$ is the entropy density. In case of thermal equilibrium in a  given DM sector 
\begin{eqnarray}
\label{Yeq}
Y_{\alpha} &=& \bar{Y}_{\alpha}= \bar{n}_{\alpha}/\mathfrak{s}(T),\\
\label{Neq}
\bar{n}_{\alpha}&=& \frac{T}{2\pi^2} \sum_{i \in \alpha} g_i m^2_i K_2(\frac{m_i}{T}).
\end{eqnarray}
Here and below we use Greek  indexes for thermal sets and Latin ones for particles. $m_i$ and $g_i$ are, respectively,  the mass  and  number of degrees of freedom of particle $i$. We will  ignore effects of Fermi and Bose statistics. Thus,  we  can calculate the rate of reaction for thermal equilibrium distributions and then rescale it for the case of real DM density. The number of events of   $a,b \to c,d$ collision  in a  space-time unit for equilibrium densities  is
\begin{equation} 
\label{nEvents22} 
          \bar{N}_{a,b \to c,d}= \frac{T g_a g_b C_{ab}}{ 8\pi^4} \int \sqrt{s}p_{ab}^2(s) K_1(\frac{\sqrt{s}}{T}) \sigma_{a,b\to c,d}(s) ds
\end{equation}
where C is a combinatoric factor:  $C_{ab}=1/2$  if $a=b$ and   1 otherwise. The thermal balance equation reads
\begin{equation}
\label{balance}
           \bar{N}_{a,b \to c,d} = \bar{N}_{c,d \to a,b} 
\end{equation}

We introduce the function  
\begin{equation}
\label{vSigmaNExp}
   \langle v\sigma_{\alpha,\beta\to\gamma,\delta}\rangle  =
\frac{1}{C_{\alpha\beta} \bar{n}_{\alpha}(T) \bar{n}_{\beta}(T)}
\sum_{\substack{a\in\alpha, b\in\beta,c\in\gamma,d\in \delta\\  if(\alpha=\beta) a\le b;\;
if(\gamma=\delta) c\le d }}  \bar{N}_{a,b \to c,d}
\end{equation}

In this notation, the  equations for the evolution of  DM abundances caused by $2\to2$ reactions read 
\begin{equation}
\label{dndt}
    \frac{dn_{\mu}}{dt} = - \sum_{\alpha\le\beta;\; \gamma \le \delta}
C_{\alpha\beta} n_{\alpha}n_{\beta}
\langle v\sigma_{\alpha,\beta\to\gamma,\delta}\rangle  ( \delta_{\mu\alpha} + \delta_{\mu\beta} - \delta_{\mu\gamma}
-\delta_{\mu\delta}) -3H(T)n_{\mu}    
\end{equation}
where $H(T)$ is the Hubble expansion rate. Taking into account entropy  conservation $\frac{d\mathfrak{s}}{dt}=-3H\mathfrak{s}$  we get 
\begin{equation}
  \label{dYdT} 
    3H \frac{dY_{\mu}}{d\mathfrak{s}} = \sum_{\alpha\le\beta;\; \gamma \le \delta}
 Y_{\alpha}Y_{\beta}
\langle v\sigma_{\alpha,\beta\to\gamma,\delta}\rangle  ( \delta_{\mu\alpha} + \delta_{\mu\beta} - \delta_{\mu\gamma}
-\delta_{\mu\delta}) 
\end{equation}

By solving these equations, the relic densities, $\Omega_{1,2,3}$, can be calculated. We rely on micrOMEGAs for numerically solving this set of coupled equations. Since its version 4.0, micrOMEGAs incorporated two-component dark matter scenarios, and has been used in the analysis of different $Z_N$ models \cite{Yaguna:2021rds, Yaguna:2021vhb, Belanger:2020hyh}. For this work, the code has been extended to models with three dark matter particles, a version that will become public shortly~\cite{micro_prep}.

Since we are dealing with a three-component dark matter scenario, it is the sum of the relic densities of each dark matter particle, $\Omega_1+\Omega_2+\Omega_3$,  that must be compared against the observed value, $\Omega_{\text{DM}}$. Here $\Omega_{\text{DM}}$ is the dark matter abundance as reported by the PLANCK~\cite{Planck:2018vyg} collaboration with an estimated theoretical uncertainty of order 10\%,
\begin{align}
    \Omega_{\text{DM}}h^2=[0.11,0.13]. 
\end{align}
As we will see, this dark matter constraint imposes severe restrictions on the viable parameter space of this model. 

 To estimate the relevance of the different kinds of processes--annihilations, semi-annihilations, and conversions--that contribute to the relic density of $\phi_1$, it is convenient to use a function in  
micrOMEGAs that allows to omit a set of processes from the relic density calculation. For example, ExcludedFor2DM="2211" will remove all conversion processes involving pairs of $\phi_2$ and $\phi_1$~\cite{Alguero:2022inz}. In particular in the following section we will compare for each component the relic density obtained from including all processes ($\Omega_i$) with the ones taking into account only processes involving DM pair annihilation into SM particles ($\Omega'_i$).
 
Another relevant quantity that must be taken into account when studying the feasibility of the model is the dark matter direct detection cross section. The elastic scattering of the dark matter particles off nuclei are possible thanks to the Higgs-portal interactions, driven by  $\lambda_{Si}$. The expression for the spin-independent (SI) cross-section then reads
 \begin{align}
     \sigma_{\phi_i}^{{\rm SI}}&=\frac{\lambda_{Si}^2}{4\pi}\frac{\mu_R^2 m_p^2 f_p^2}{m_h^4 M_{i}^2},
 \end{align}
 where $\mu_R$ is the reduced mass, $m_p$ the proton mass and $f_p\approx 0.3$ is the quark content of the proton. Given that we have multiple dark matter particles, the  quantity to be compared against the direct detection limits provided by the experimental collaborations is not the cross section itself but rather the products $\xi_i \sigma_{\phi_i}^{{\rm SI}}$ with $\xi_i\equiv \Omega_i/\Omega_{\text{DM}}$. 
To impose current limits from the LZ experiment~\cite{LZ:2022ufs}, we use 
micrOMEGAs which  calculates  the 90\% confidence level exclusion  in DD  experiments  using  the recoil   energy distribution  of nuclei~\cite{Belanger:2020gnr,micro_prep}.  
For multicomponent DM model, the total energy distribution is calculated where the contribution of each component is weighted by the factor $\xi_i$. 
Note that we have explicitly checked that in all cases where the DM components have similar masses this procedure gives a similar result than the one obtained by simply comparing $\sum_i \xi_i \sigma_{\phi_i}^{SI}$ with the 90\% exclusion limit.
In this work we also consider,  for each of the dark matter particles,  the detection prospects in future experiments, in particular DARWIN~\cite{Aalbers:2016jon}.

\section{The viable regions}
\label{sec:viable}
In this section we present our main results: the viable regions of the $Z_7$ model are found and characterized, and its detection prospects are analyzed. To that end, we randomly scan the parameter space of the model in different ways, so as to identify the most relevant effects and the parameters that determine them. From each of these scans, we obtain a large sample of viable models. A  model is considered viable if it is consistent with all known experimental constraints, including the relic density and the LZ direct detection limit. The viable models are then projected onto different planes for each random scan, and their implications separately discussed.

Among the free parameters of this model, the dark matter-Higgs couplings ($\lambda_{Si}$) are the most relevant because they link the SM and the DM sectors, allowing the dark matter particles to reach thermal equilibrium in the early Universe --a prerequisite for freeze-out. In addition, these same couplings determine the dark matter direct detection cross sections,  which provide the main avenue to test this scenario. Consequently, in all our scans the couplings $\lambda_{Si}$ will be taken to be different from zero. In fact they vary within the range: 
\begin{align}
    10^{-4} \leq |\lambda_{Si}| \leq1.
\end{align}

All other couplings could in principle vanish. If they all vanished simultaneously, however, the model would become equivalent to three copies of the complex singlet scalar model, which is known to be viable only for heavy DM.
Indeed an upper limit on the  coupling $\lambda_{Si}$ is derived from direct detection and that coupling cannot provide efficient DM annihilation in the early Universe unless  the DM mass is around 6 TeV. 
That case is of no interest to us. What we want to study is precisely if and how the new couplings present in the $Z_7$ model allow to satisfy the cosmological and experimental bounds for masses smaller than those in the singlet scalar case.  Since $\phi_1$ is the lightest dark matter particle ($M_1<M_2<M_3$), these new couplings must facilitate the annihilation of $\phi_1$ in the early Universe.  And we already know, from the study of previous $Z_N$ models, that the trilinear couplings are more effective than the quartic ones in opening up new regions of parameters space. Taking these facts into consideration, we have done a series of scans in  which only a subset of the free parameters of the model are different from zero at a time. In this way we expect to identify the most relevant couplings as well as the new processes that allow to satisfy the experimental constraints.  The four different scans we will analyze in detail are:

\begin{enumerate}
    \item[A)] $M_i\neq 0, \lambda_{Si}\neq0, \mu_{S4}\neq0$, 
    \item[B)] $M_i\neq 0, \lambda_{Si}\neq0, \mu_{S1}\neq0, \mu_{S2}\neq0, \mu_{S3}\neq0$,
    \item[C)] $M_i\neq 0, \lambda_{Si}\neq0,  \mu_{S1}\neq0, \lambda_{4ij}\neq0$, 
    \item[D)] $M_i\neq 0, \lambda_{Si}\neq0,  \mu_{S3}\neq0, \lambda_{4ij}\neq0, \lambda_{34}\neq0$,
\end{enumerate}
where, in each case, the rest of free parameters has been set to zero. For instance, in scan C) the parameters $\mu_{S4}, \lambda_{4ij}, \lambda_{3i}$ were all set to zero. In all the scans, the parameters that are different from zero are allowed to randomly vary within the following ranges: 
\begin{align}
    50\, {\rm GeV} \leq &M_1 \leq 2\, {\rm TeV},\nonumber\\
    M_1 <&M_2<2M_1,\nonumber\\
    M_2<&M_3<M_1+M_2,\nonumber\\
    10^{-4}\leq &|\lambda_{4ij}| \leq1,\nonumber\\
    100\, {\rm GeV} \leq &\mu_{Si} \leq 10\, {\rm TeV}.
\end{align}
Next, the results for each of the scans A-D are separately presented. Since $M_i,\lambda_{Si}\neq 0$ is common to all, we will label them according to the other couplings that are not zero.

\subsection{$ \mu_{S4}\neq0$}
This scenario  presents semiannihilation processes that involve the three dark matter particles,
\begin{align}
 \phi_i+\phi_j\to \phi_k^\dagger +h, \hspace{1cm}   i\neq j \neq k,
\end{align}
with $\phi_k$ ($\phi_j$) acting as the mediator in the $s$ ($t$) channel. These processes  have no counterpart within two-component dark matter scenarios and their impact on the relic densities has not been investigated before. Notice that the number density of the two incoming particles is reduced by one unit while the number density of the outgoing particle is increased by one unit. In this way,  the three relic densities can be affected by a  single trilinear interaction. 

\begin{figure}
\centering
\includegraphics[scale=0.4]{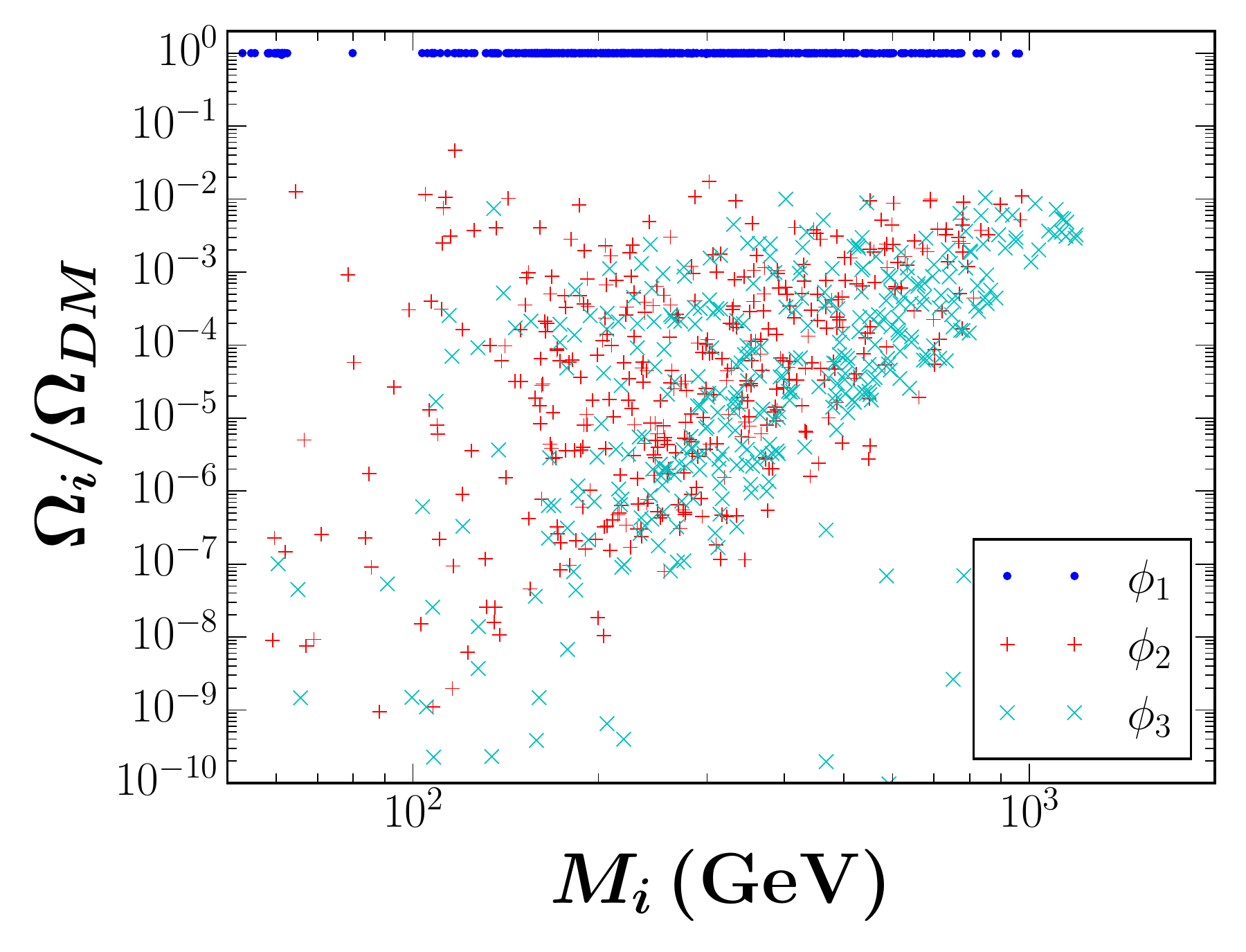}\hspace{1cm}
\includegraphics[scale=0.4]{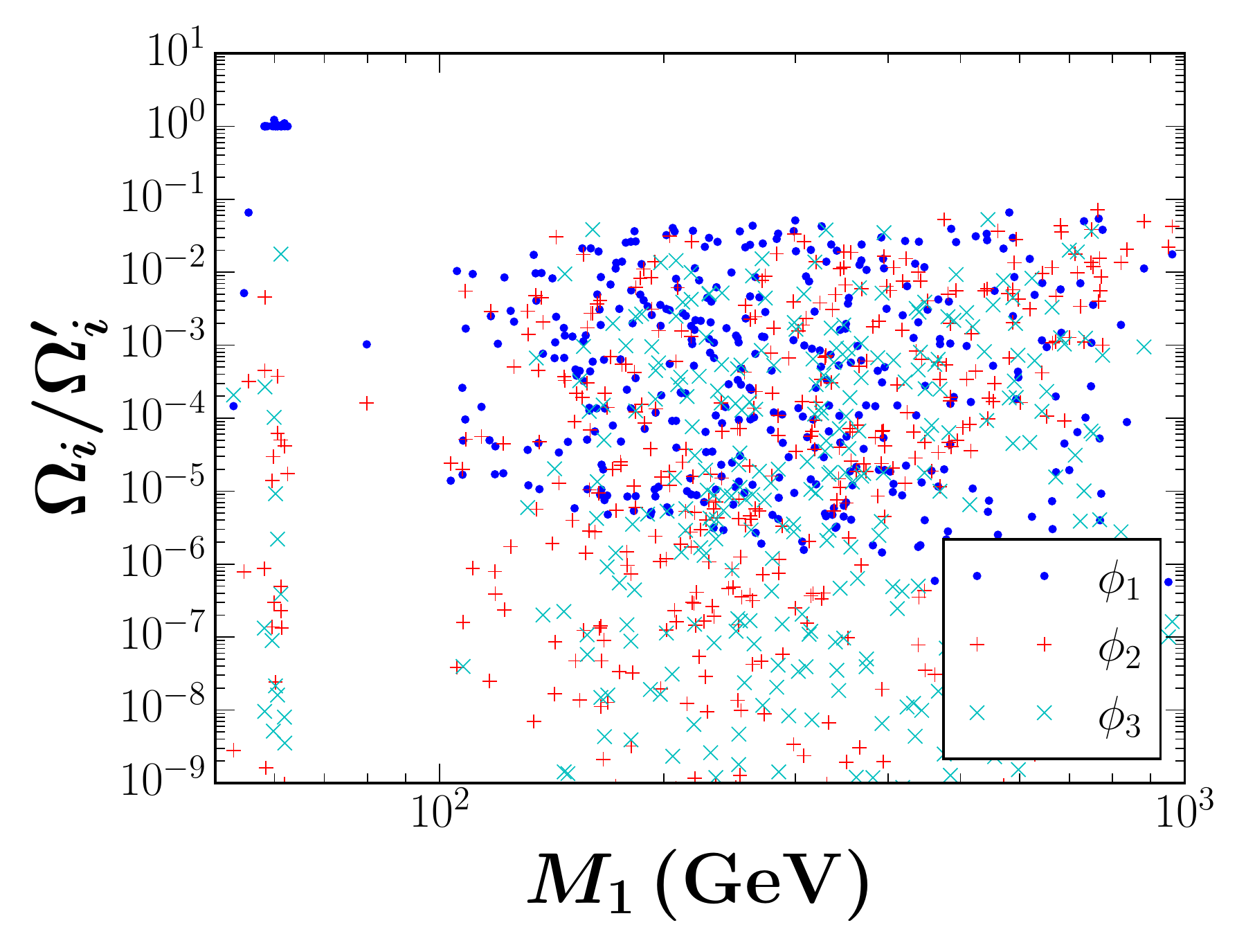}\\
\includegraphics[scale=0.4]{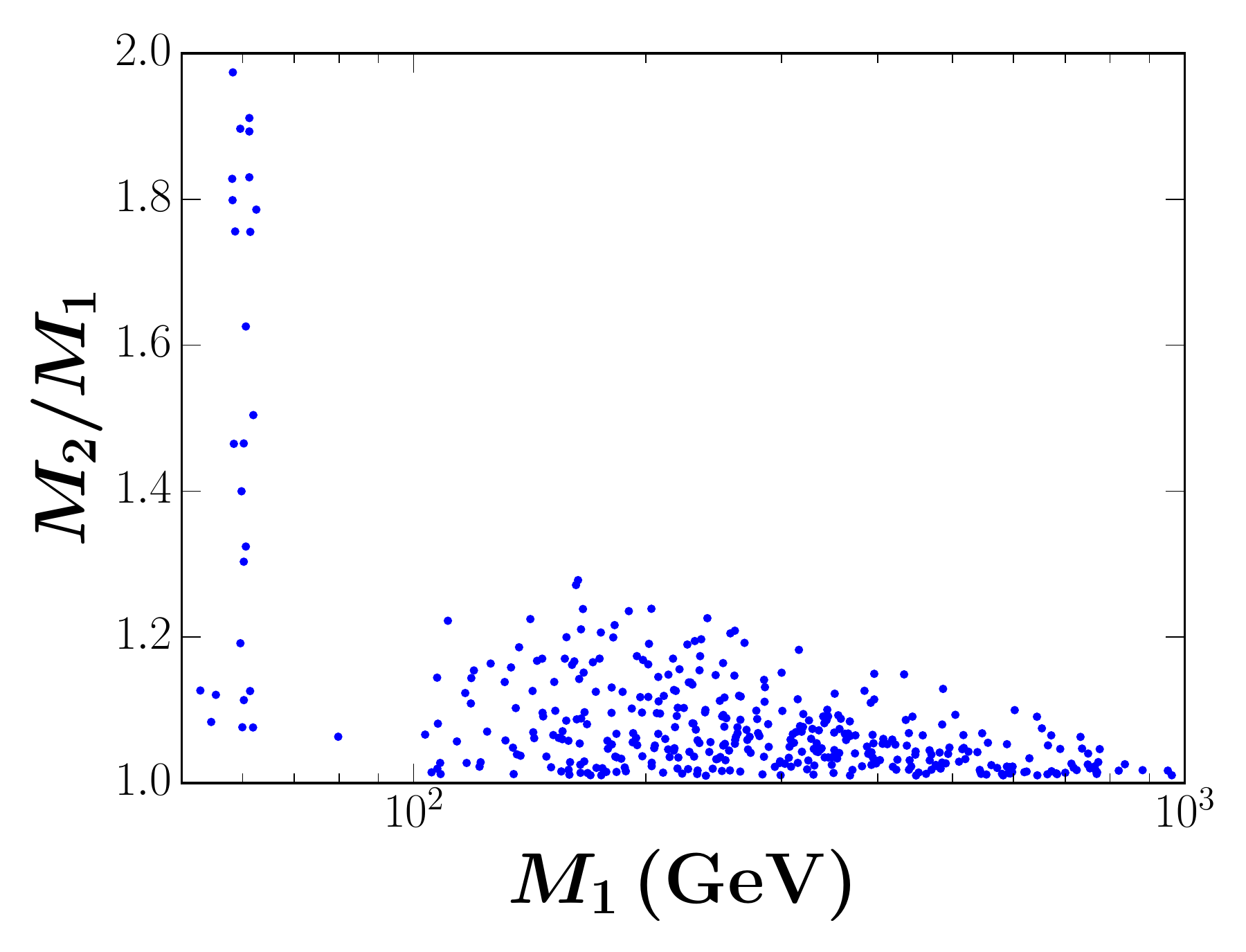}\hspace{1cm}
\includegraphics[scale=0.4]{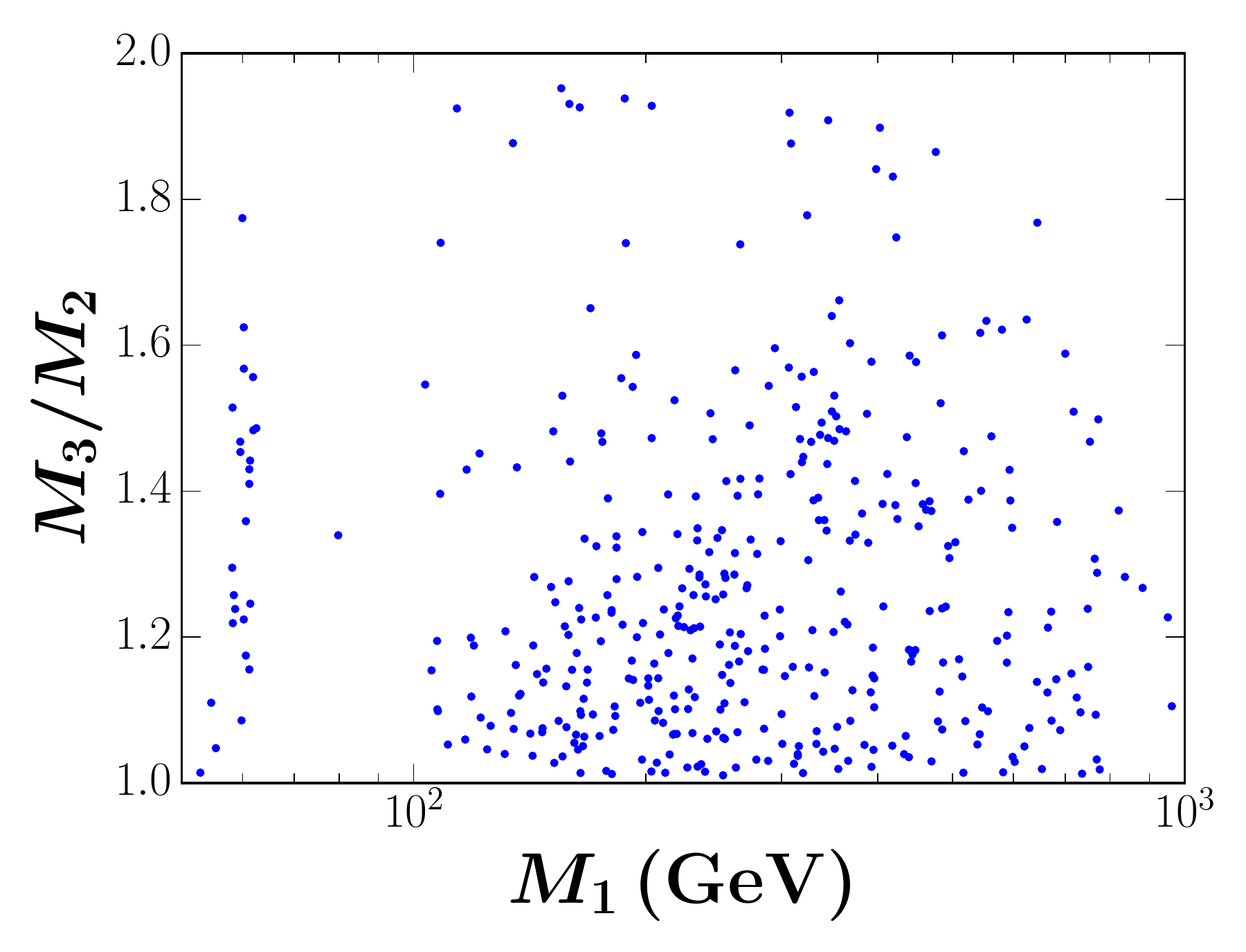}\\
\includegraphics[scale=0.4]{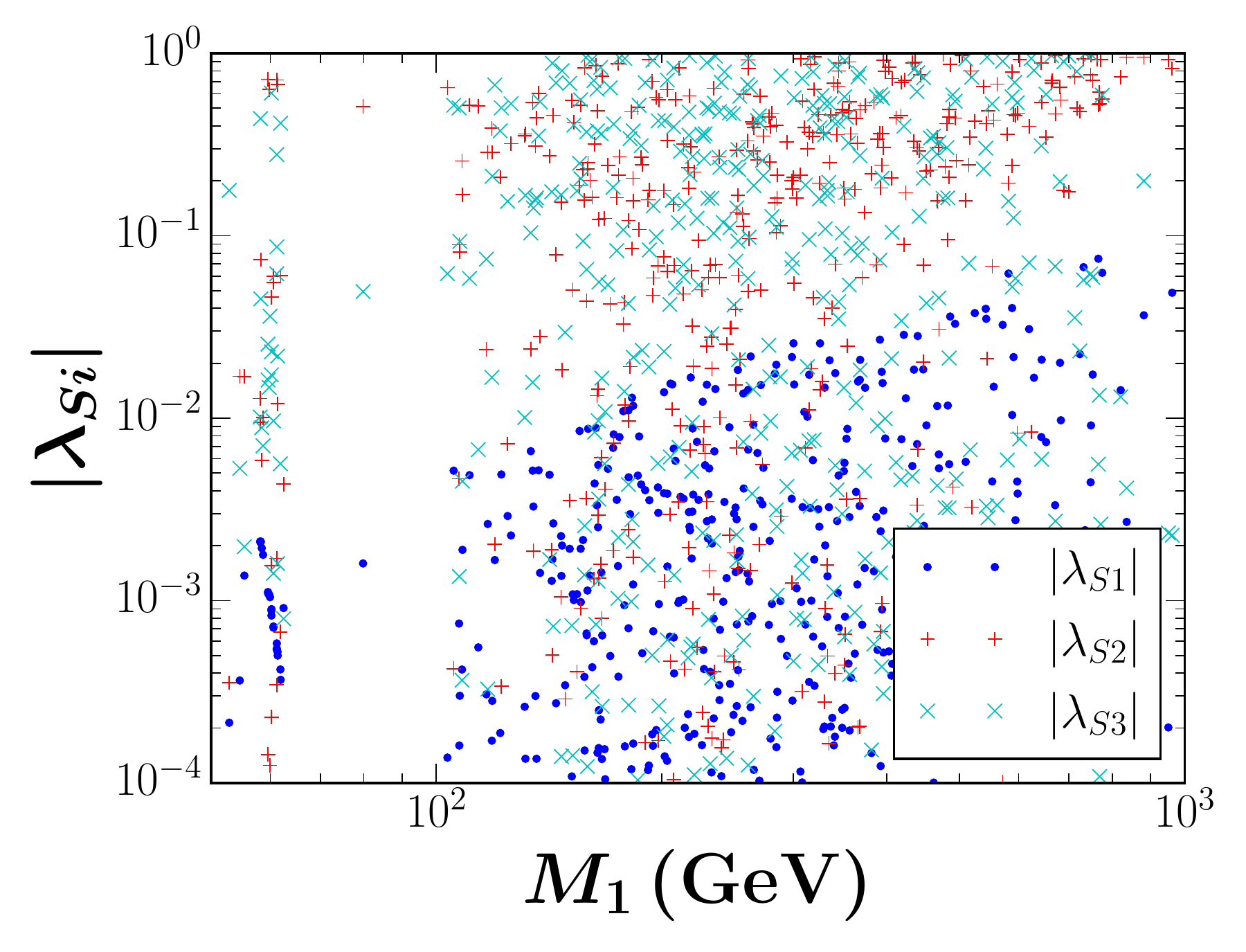}\hspace{1cm}
\includegraphics[scale=0.4]{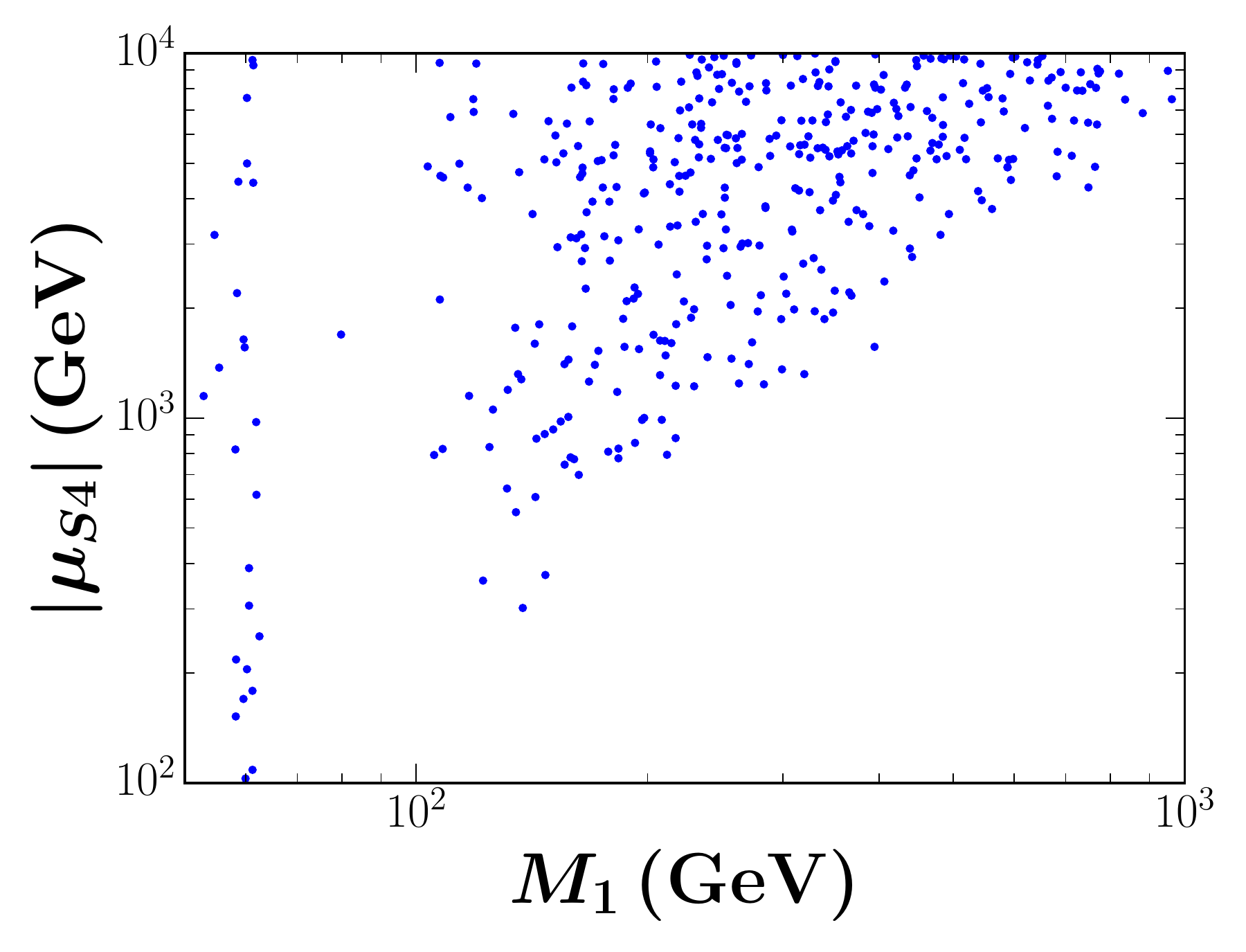}
\caption{The viable regions for $\mu_{S4}\neq0$ (scan A). All points shown in these plots are compatible with current data. }
\label{fig:scan-mus4}
\end{figure}

\begin{figure}
\centering
\includegraphics[scale=0.5]{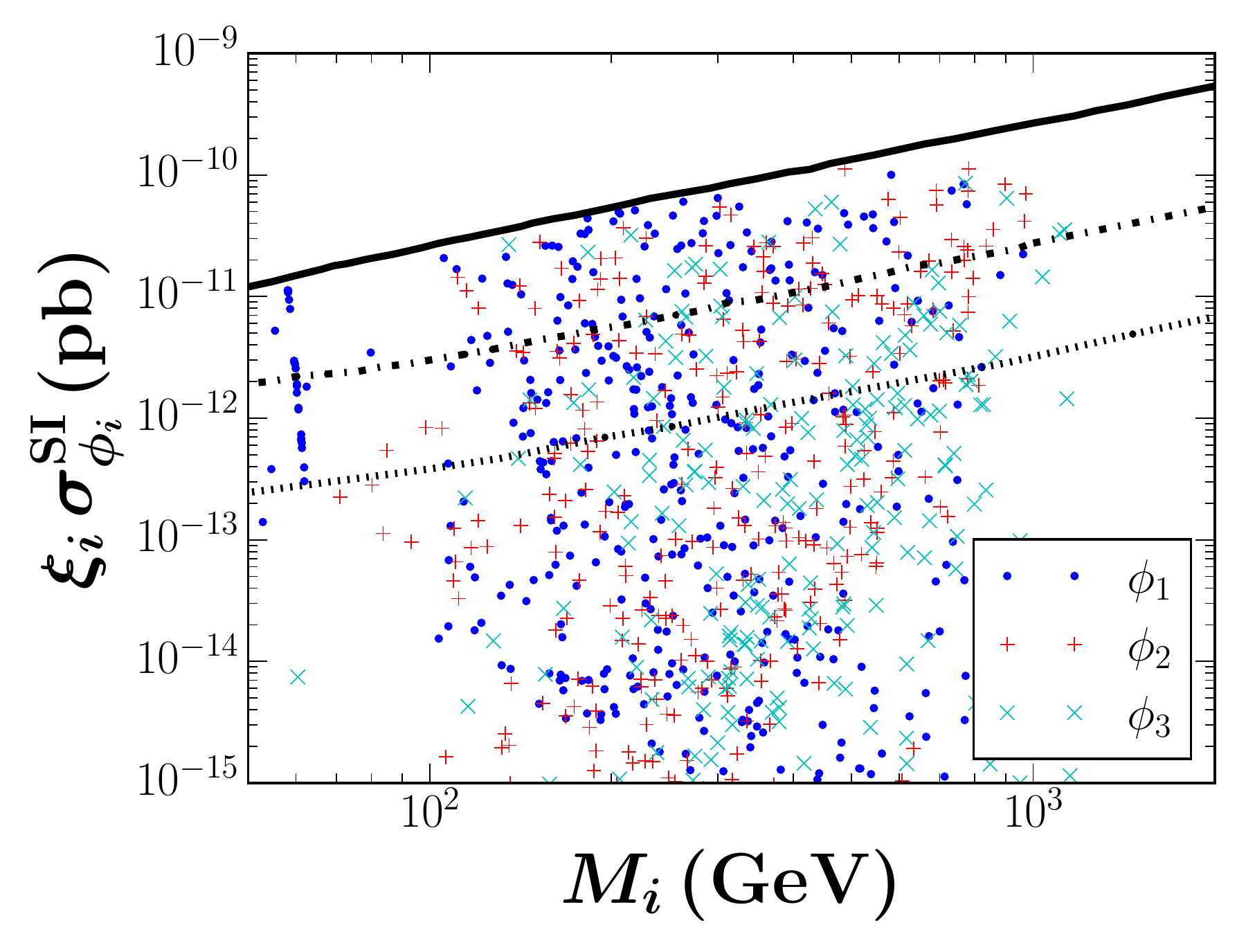}

\caption{Detection prospects in the scan with $\mu_{S4}\neq0$ (scan A). The solid line is the current LZ limit whereas the dash-dotted and  dotted lines correspond to the expected sensitivities of LZ~\cite{LZ:2022ufs} and DARWIN~\cite{Aalbers:2016jon} respectively.}
\label{fig:scan-mus4det}
\end{figure}

Figure \ref{fig:scan-mus4} shows the set of viable models projected onto different planes. Notice, first of all, that the viable models span a wide range of dark matter masses, with the lightest ($M_1$) extending up to $1$ TeV and the heaviest ($M_3$) up to $2$ TeV --a region that is excluded in the (complex) singlet scalar model. Thus, the new interactions allowed by the $Z_7$ symmetry are indeed opening up new regions of viable parameter space at low masses. From the top-left panel we learn that $\phi_1$ (the lightest dark matter particle) always gives the dominant contribution to the dark matter density, with $\phi_2$ and $\phi_3$ typically contributing less than $1\%$.  The top-right panel indicates that the relic density of all components is strongly reduced when including dark matter conversions and semiannihilations except when $M_1\approx m_h/2$ and annihilation is enhanced by the Higgs resonance. For $\phi_1$ the relic density is determined by conversions, $\phi_1\phi_1^\dagger \to \phi_2\phi_2^\dagger$, and semiannihilations, $\phi_1\phi_2 \to \phi_3^\dagger h$. The former requires a small mass splitting $M_2-M_1$  while the latter is relevant for moderate values of $M_3/M_2$. This hierarchy among the dark matter masses is illustrated on the next row. 
From the left panel, we see that outside the Higgs resonance $\phi_1$ and $\phi_2$ have similar masses, a tendency that becomes stronger at higher masses. The ratio $M_3/M_2$, on the contrary, vary over the entire allowed range --see right panel. The bottom-left panel displays the coupling of the Higgs to the dark matter particles. Due to the direct detection constraints, $\lambda_{S1}$ is always  small. On the other hand  $\lambda_{S2,S3}$ can be ${\cal O}(1)$ since the contributions of $\phi_2$ and $\phi_3$ to DM scattering on nucleons is suppressed by  $\xi_2,\xi_3 < 10^{-2}$.  The other relevant coupling is $\mu_{S4}$, which, giving its role in setting the $\phi_1$ relic density, tends to increase with $M_1$ --see bottom-right panel.

The detection  prospects for this scenario are illustrated in figure \ref{fig:scan-mus4det}, which compares the elastic scattering cross section for each of the dark matter particles against the current limit from LZ (solid line) and the expected sensitivity of future experiments (dotted lines).  Remarkably, we find that all the dark matter particles might give rise to observable signals in future experiments. That is, even $\phi_2$ and $\phi_3$, whose contributions to the dark matter density are below $1\%$ (see previous figure), could be detected. What explains this counterintuitive result is the fact that the smaller dark matter fraction is compensated by larger Higgs-portal couplings --see bottom-left panel in figure \ref{fig:scan-mus4}. Note, in particular, that a significant fraction of the viable models in our scan lies within the expected sensitivity of DARWIN. Direct detection, therefore, constitutes a promising way of probing this scenario.

\subsection{$\mu_{S1}\neq0, \mu_{S2}\neq0,\mu_{S3}\neq0 $}

\begin{figure}
\centering
\includegraphics[scale=0.4]{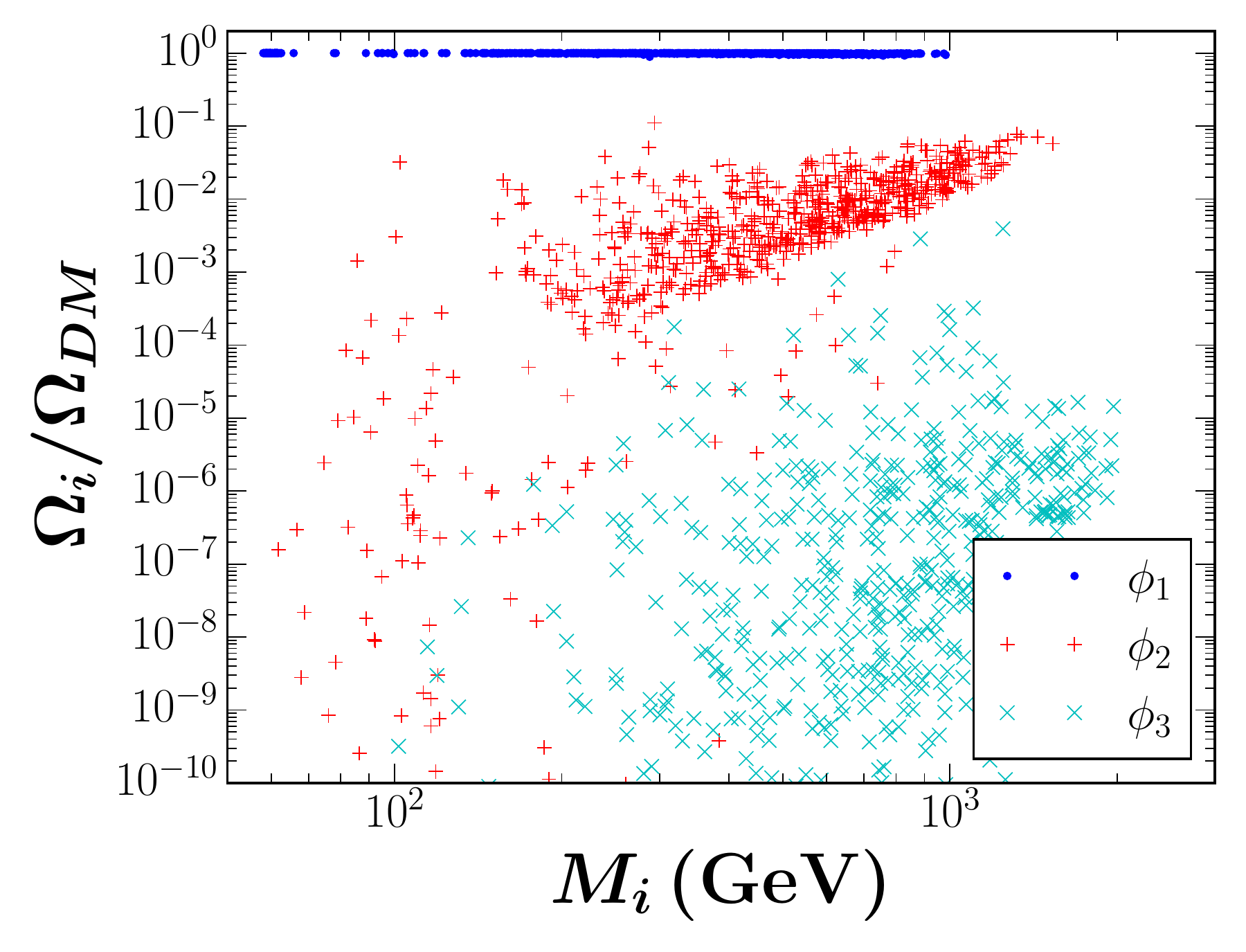}\hspace{1cm}
\includegraphics[scale=0.4]{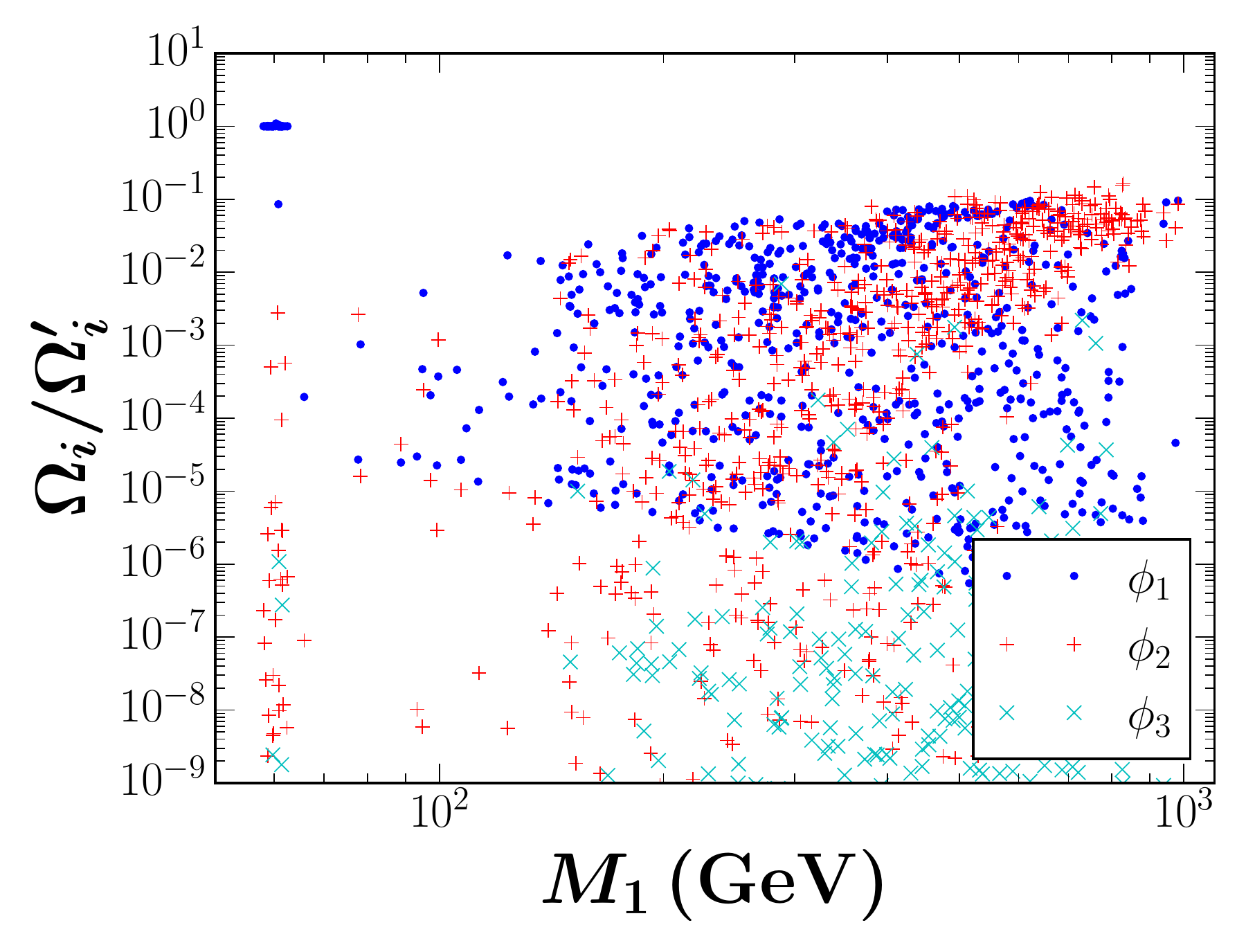}\\
\includegraphics[scale=0.4]{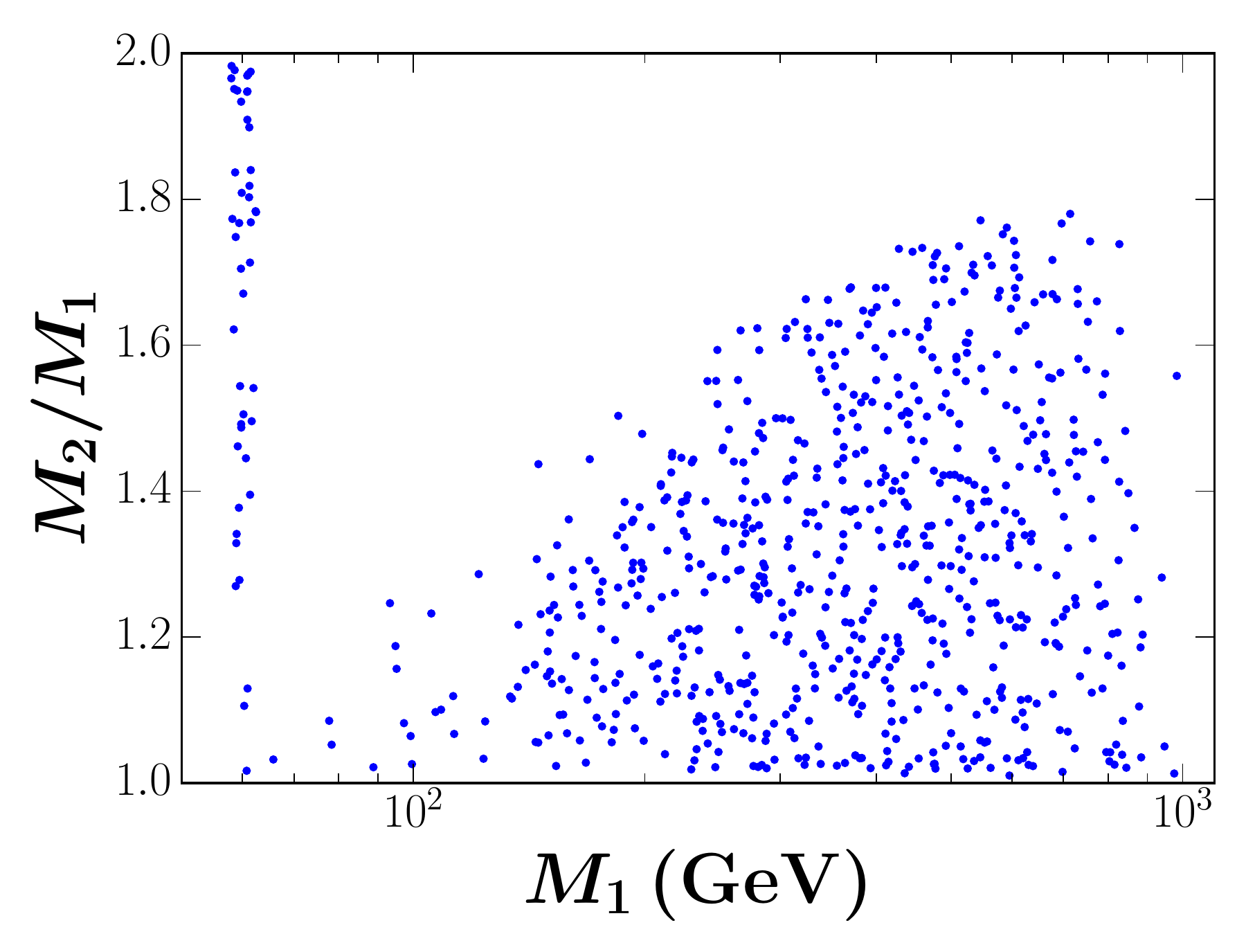}\hspace{1cm}
\includegraphics[scale=0.4]{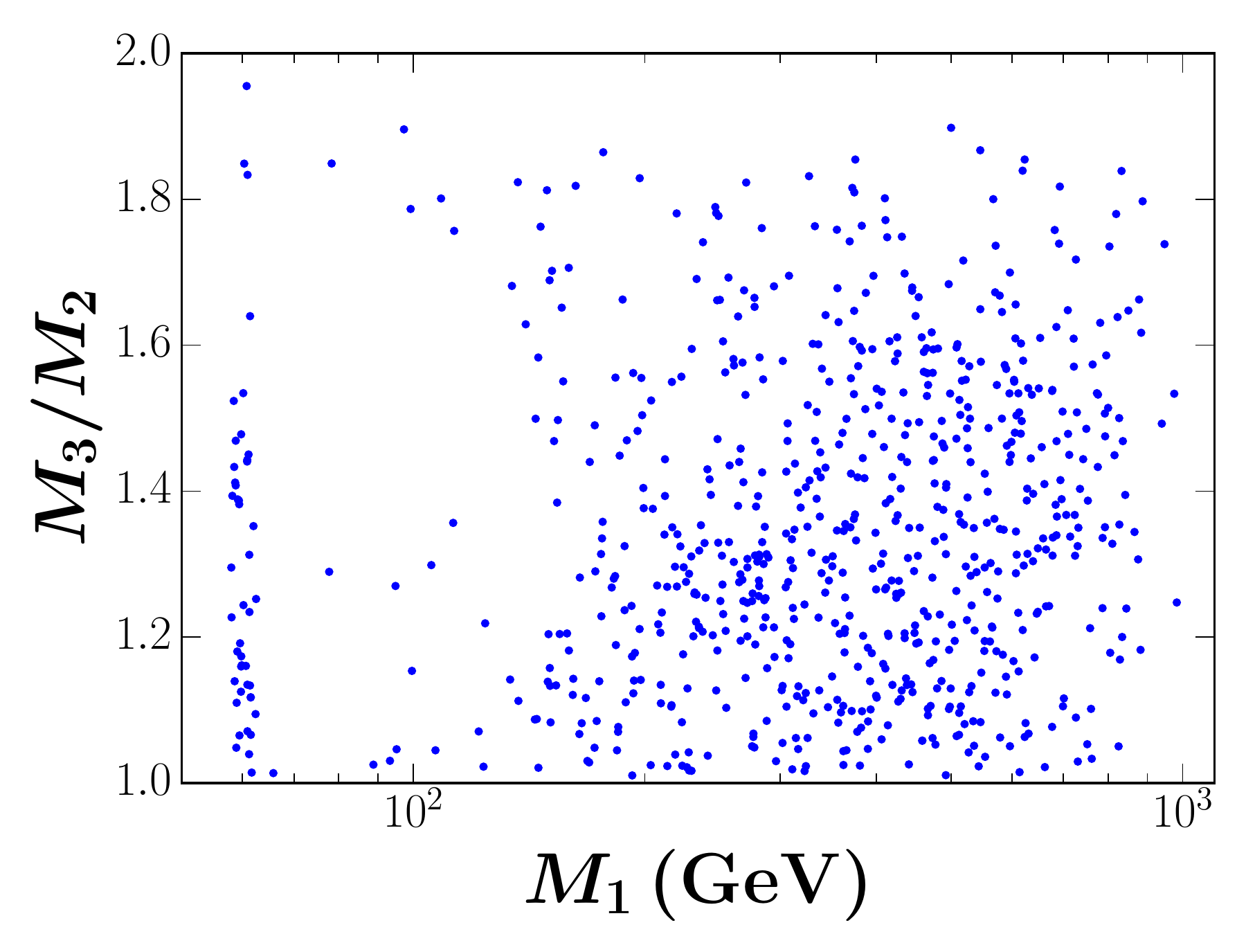}\\
\includegraphics[scale=0.4]{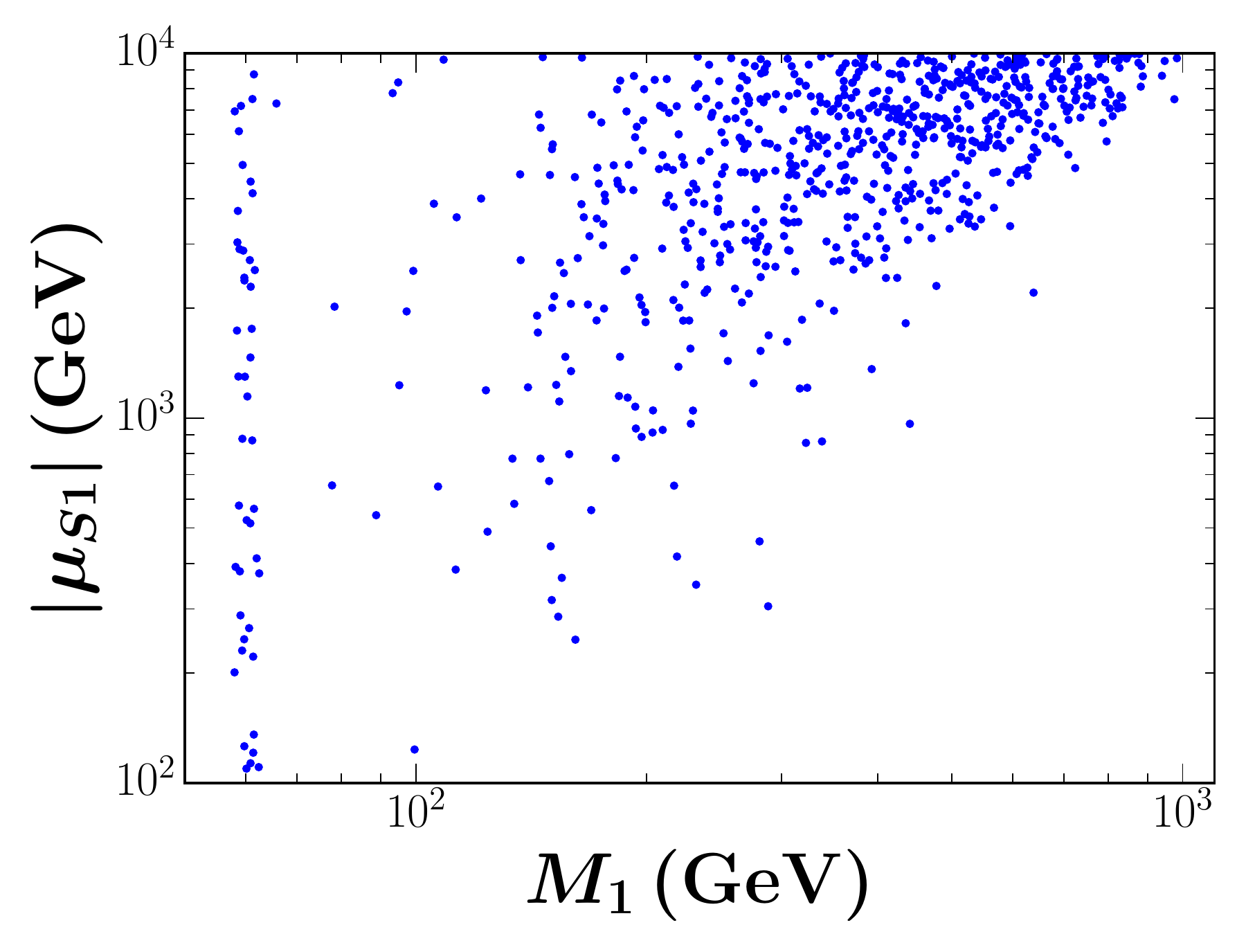}\hspace{1cm}
\includegraphics[scale=0.4]{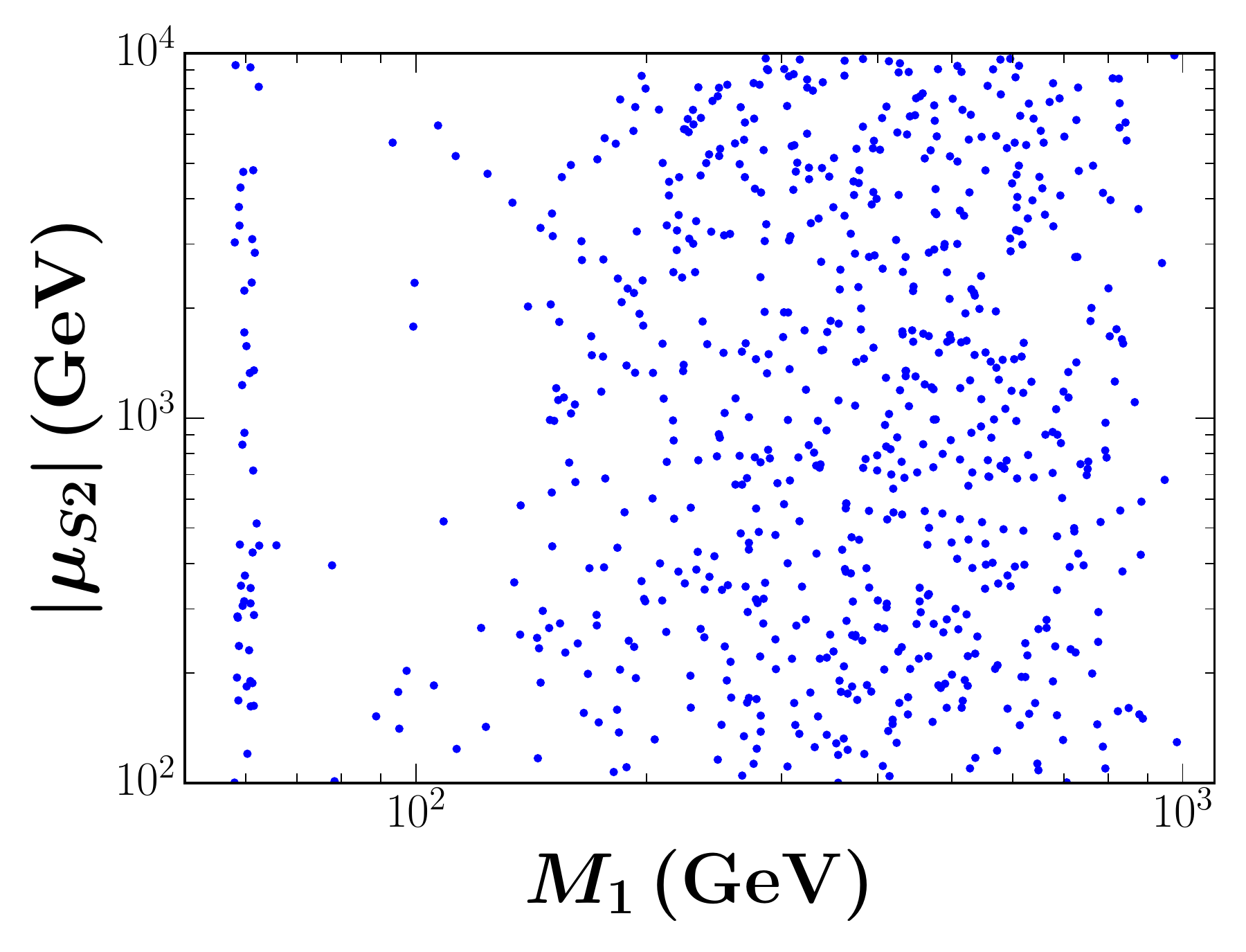}\\
\includegraphics[scale=0.4]{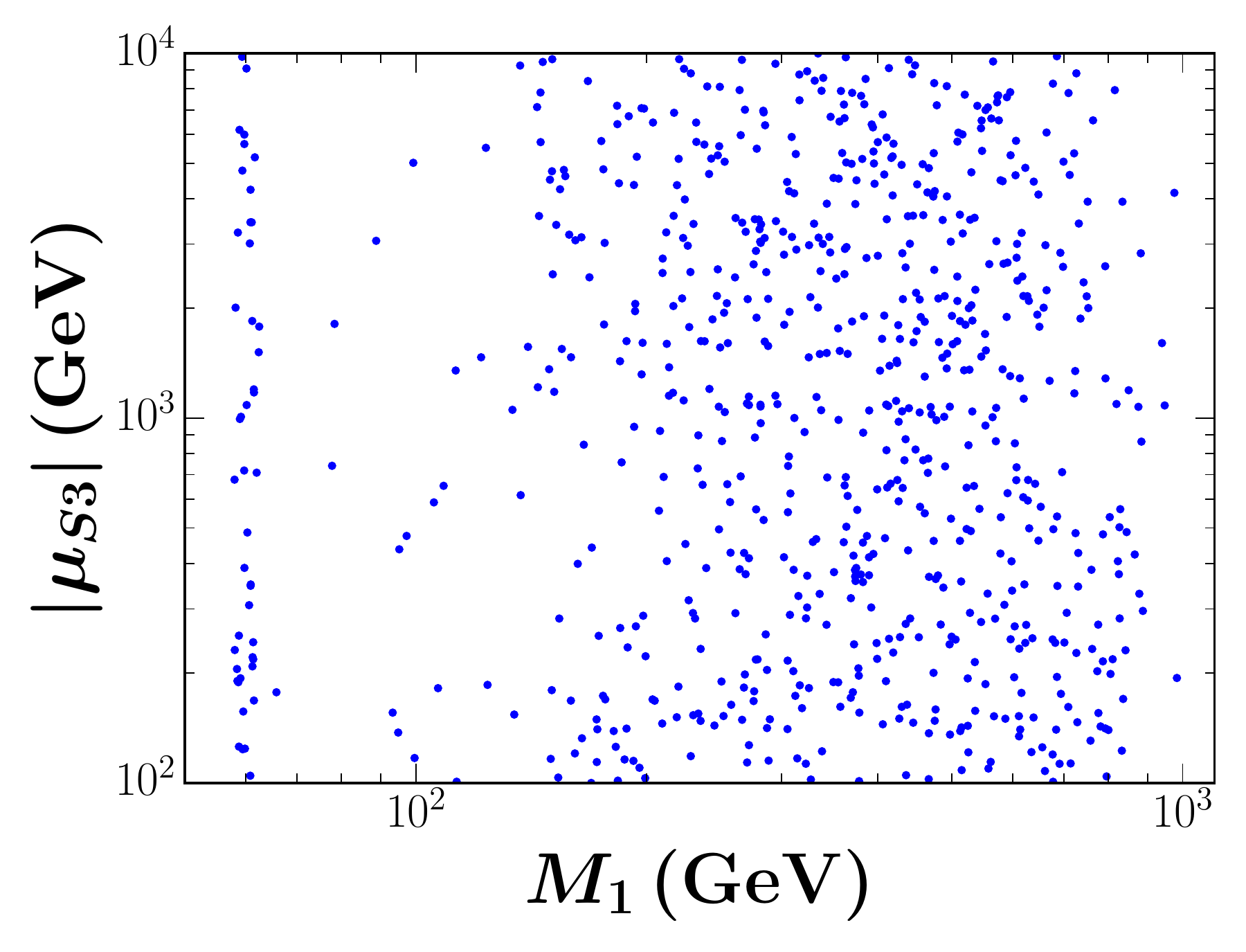}\hspace{1cm}
\includegraphics[scale=0.4]{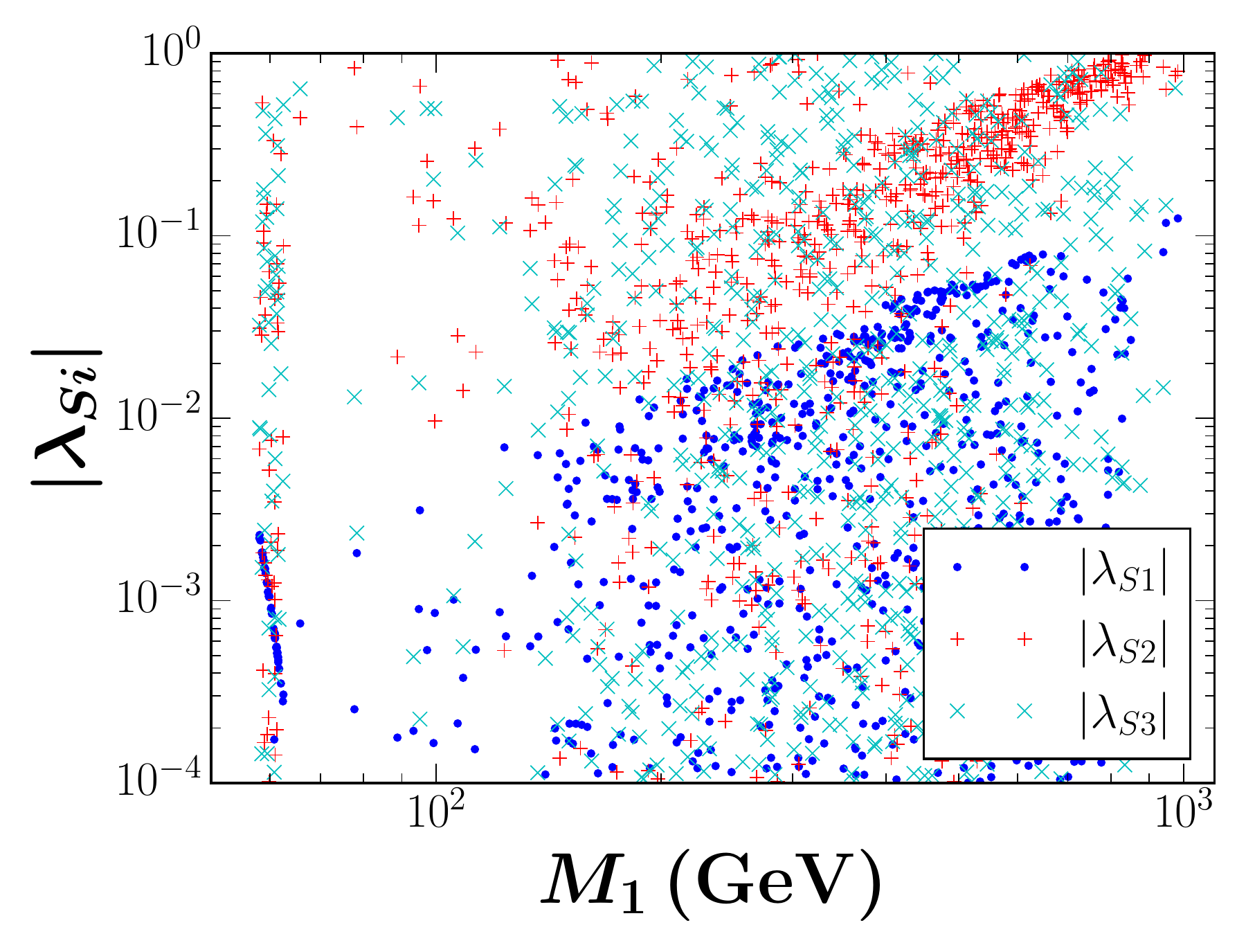}
\caption{The viable regions for $\mu_{S1}\neq0$, $\mu_{S2}\neq0$, $\mu_{S3}\neq0$ (scan B). All points shown in these plots are compatible with current data.}
\label{fig:scan-mus1-mus2-mus3}
\end{figure}

For our second scan we allow the trilinear terms $\mu_{S1}$,  $\mu_{S2}$ and $\mu_{S3}$ to simultaneously vary. The main novelty with respect to the previous scenario is the appearance of dark matter conversion processes involving two different trilinear couplings such as  the s-channel process $\phi_1+\phi_3\to \phi_2+\phi_2$ mediated by $\phi_3$ --another example of a process involving three different dark matter particles.  Conversion processes of the type 1122  are also relevant and  they may all affect the relic densities of the three dark matter particles. 

The set of viable models for this case is displayed in figure  \ref{fig:scan-mus1-mus2-mus3}.  Regarding the relic density, we learn from the top-left panel  that it is always dominated by $\phi_1$, with $\phi_2$ giving at most a $10\%$ contribution, and a much smaller one from $\phi_3$. As in the previous case, conversion and semiannihilation processes provide the dominant contribution to the relic density of the three components, as shown in the top-right panel. For example for $\Omega_1$,  conversion processes involving either $\phi_2$ and $\phi_3$ play a role while as before, $\phi_1\phi_1 \to \phi_2 h$ is the relevant  semiannihilation process, it reduces the density of $\phi_1$ when $M_2+m_h< 2 M_1$. This process is more important for large values of $\mu_{S1}$. The next row demonstrates that, for our sample of viable models, the dark matter masses vary pretty much over their entire allowed range. Thus, in contrast with the previous scan, no degeneracies are required in this case. The next two rows display the trilinear and the Higgs-portal couplings. Notice that $\mu_{S1}$ tends to increase with $M_1$ whereas $\mu_{S2}$ and $\mu_{S3}$ vary within their allowed range without a clear pattern. From the bottom-right panel, we see that, as a result of the direct detection constraint, $\lambda_{S1}$ (blue points) is suppressed, with a maximum value below $0.1$, while $\lambda_{S2}$ (red points) tends to be large, foreseeing a significant direct detection rate. 

\begin{figure}
\centering
\includegraphics[scale=0.5]{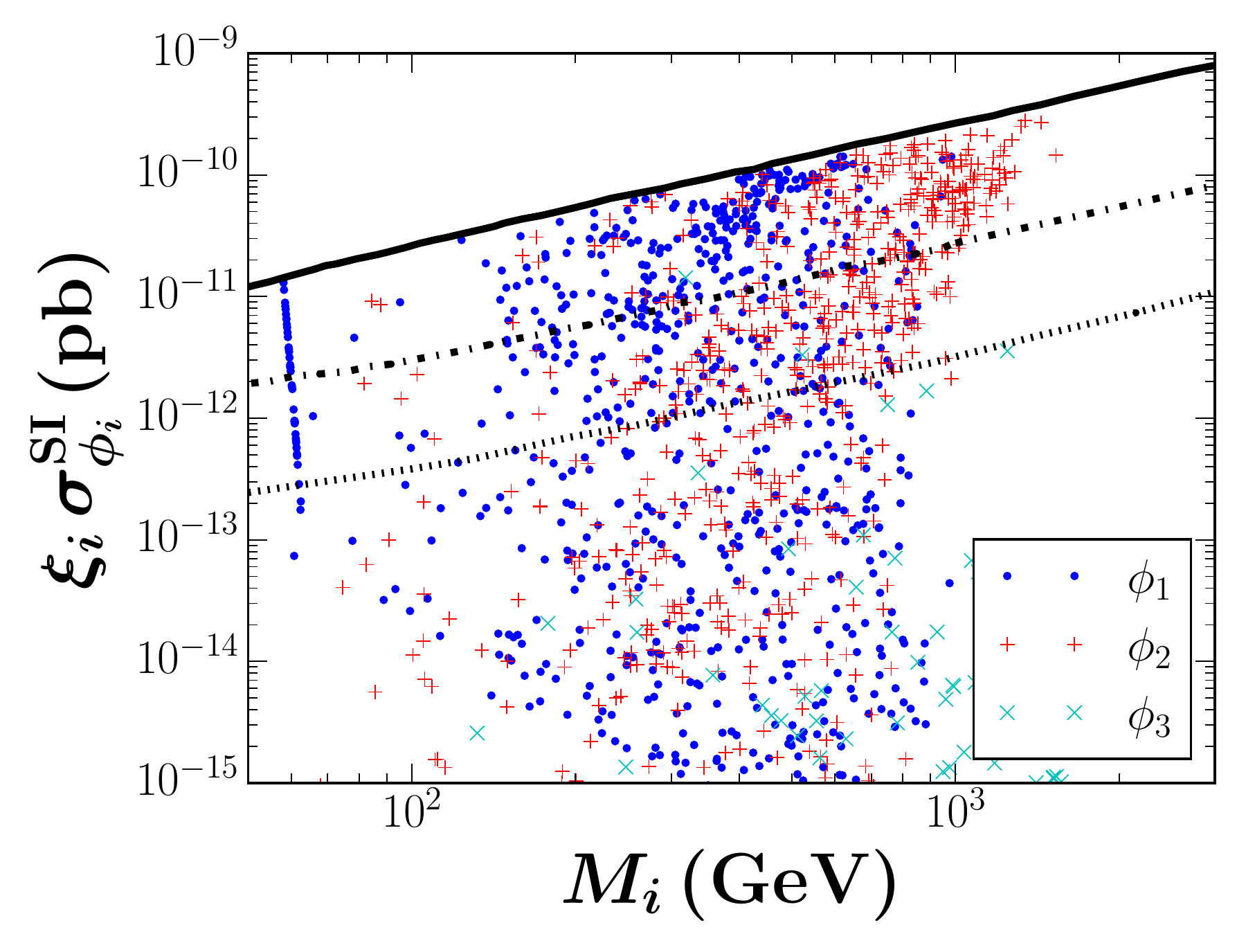}
\caption{Detection prospects in the scan with $\mu_{S1},\mu_{S2}, \mu_{S3}\neq0$ (B). The solid line is the current LZ limit whereas the dash-dotted and  dotted  lines correspond to the expected sensitivities of LZ and DARWIN respectively}.
\label{fig:scan-mus123det}
\end{figure}

Figure \ref{fig:scan-mus123det} shows the rescaled direct detection cross sections for each of the dark matter particles, and compares them against current data (solid line) and future sensitivities (dotted lines). Notice that while $\phi_1$ and $\phi_2$ often give rise to signals within the expected sensitivy  of DARWIN, $\phi_3$ rarely does so. Thus, in this scenario, one would expect to observe at most two dark matter particles in future direct detection experiments.  This may allow to exclude the standard paradigm with just one dark matter particle, but it would be more challenging to establish that it corresponds to a three-component dark matter scenario. In any case, direct detection provides a robust way to test this possibility.

\subsection{$\mu_{S1}\neq0, \lambda_{4ij}\neq0$}
\begin{figure}
\centering
\includegraphics[scale=0.4]{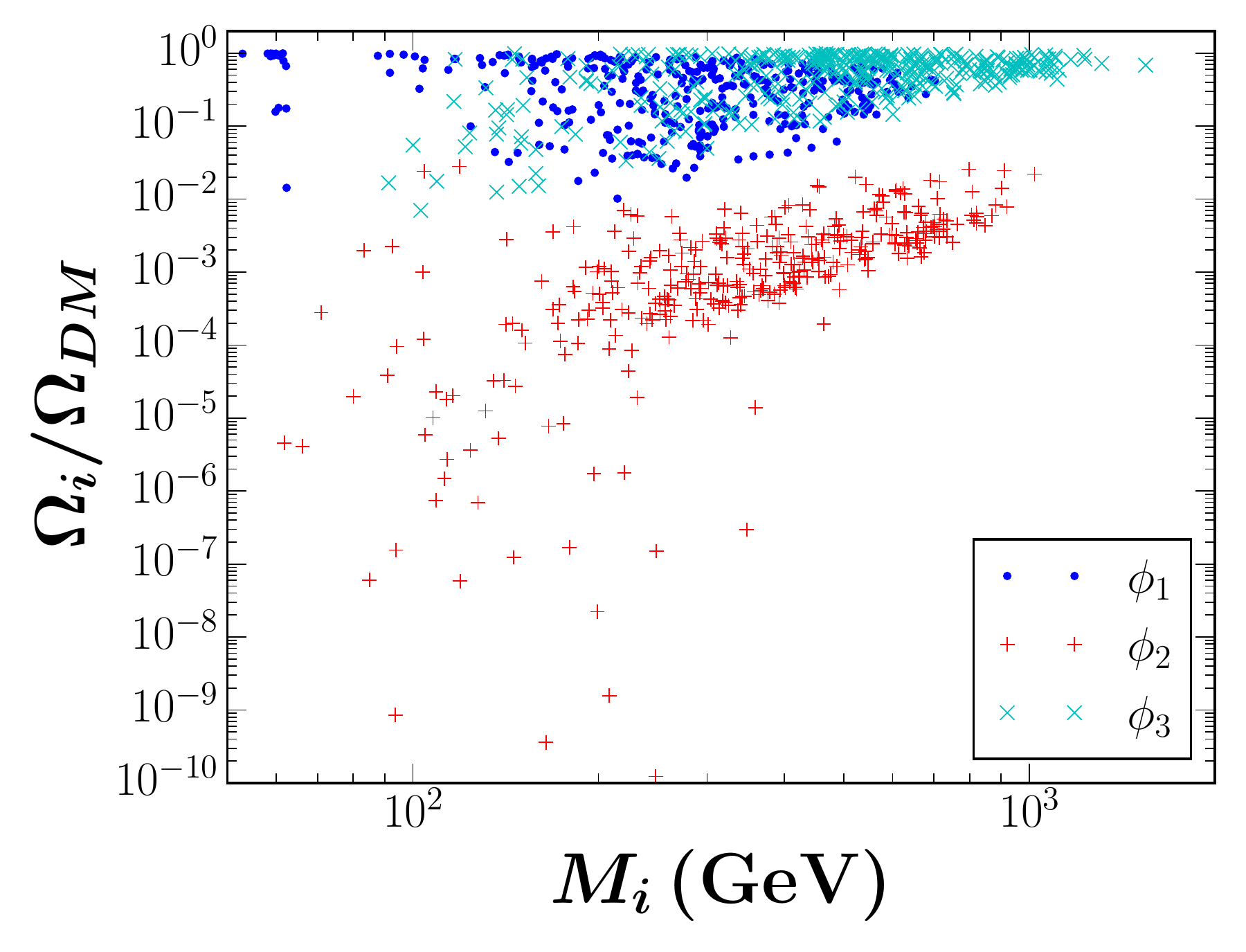}\hspace{1cm}
\includegraphics[scale=0.4]{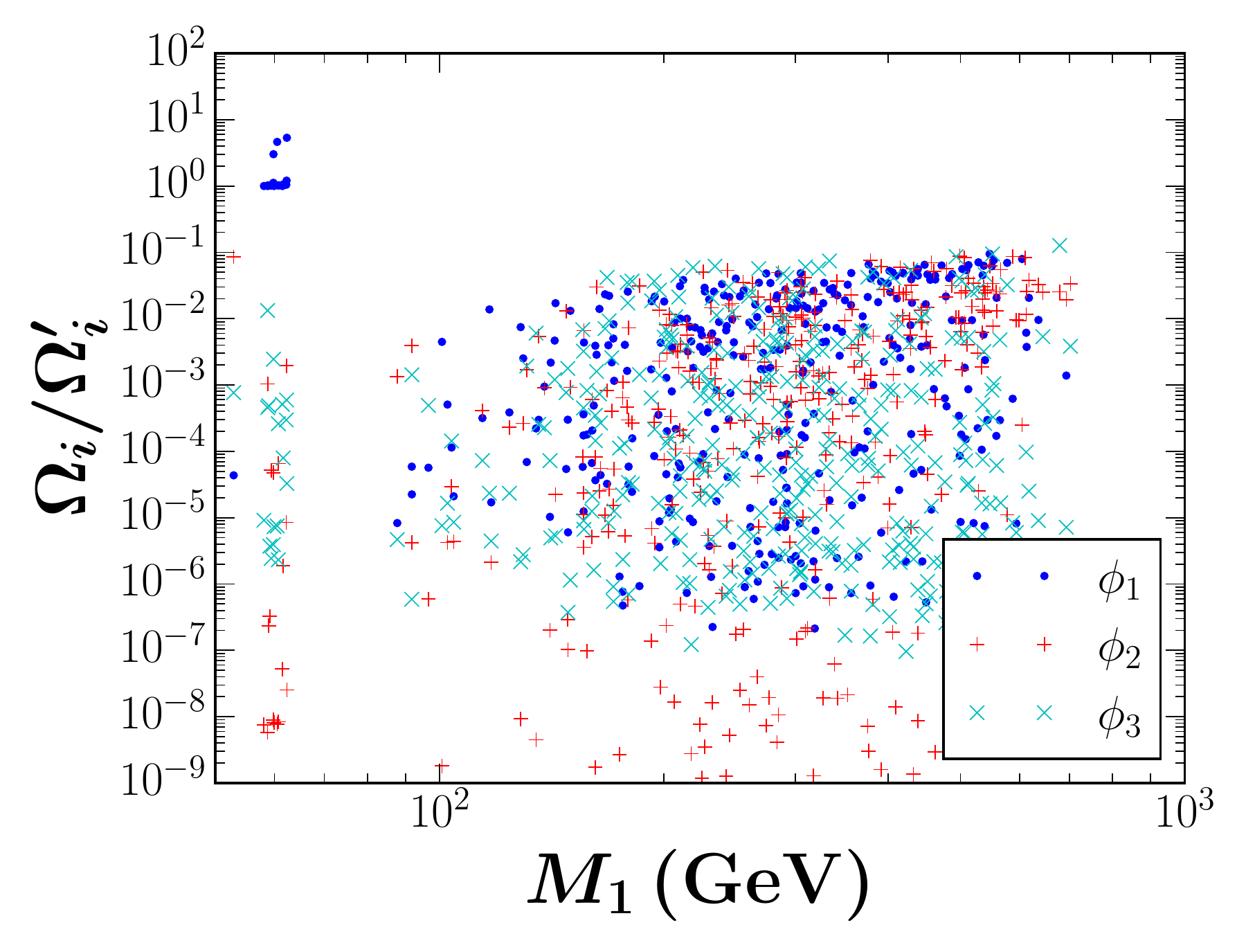}\\
\includegraphics[scale=0.4]{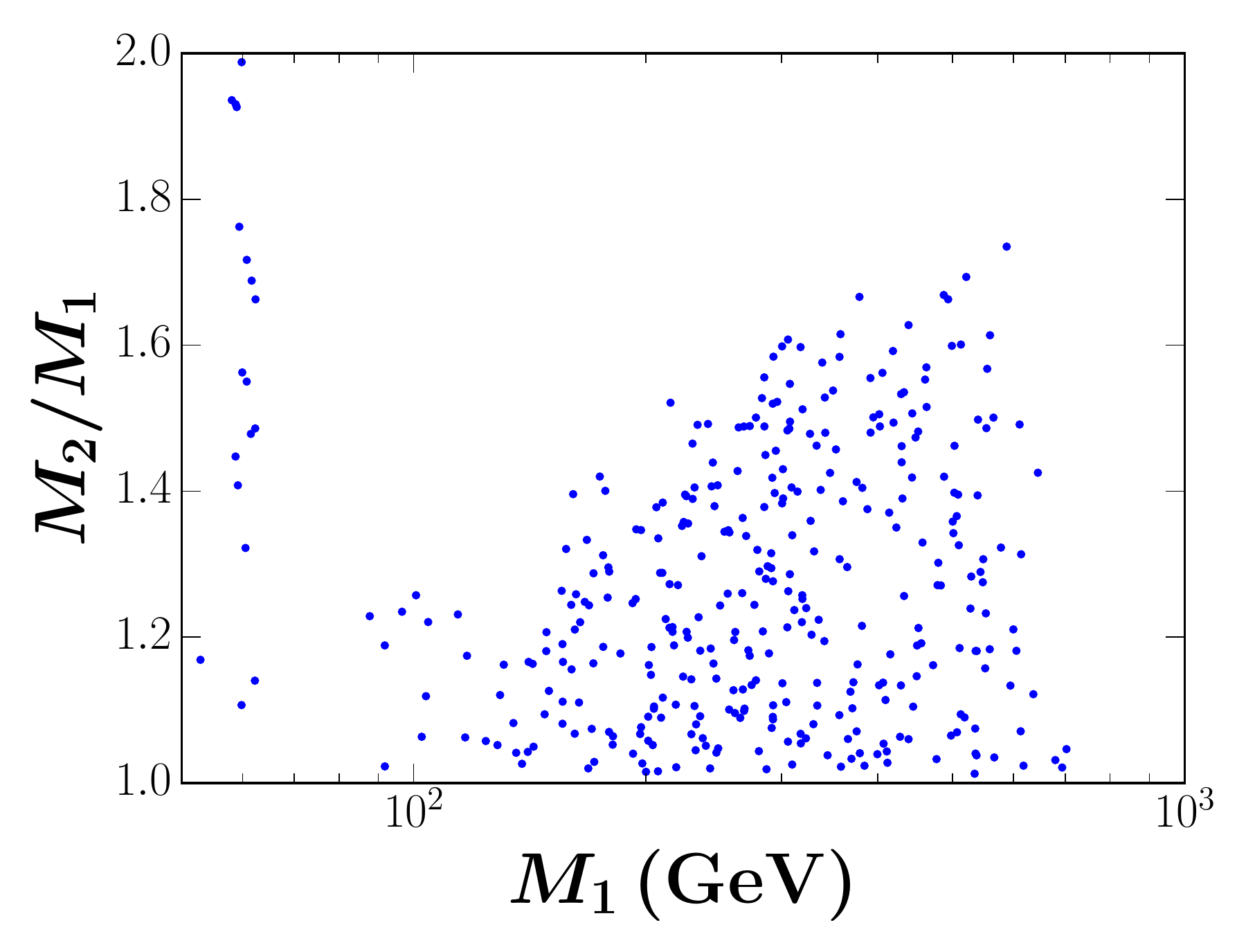}\hspace{1cm}
\includegraphics[scale=0.4]{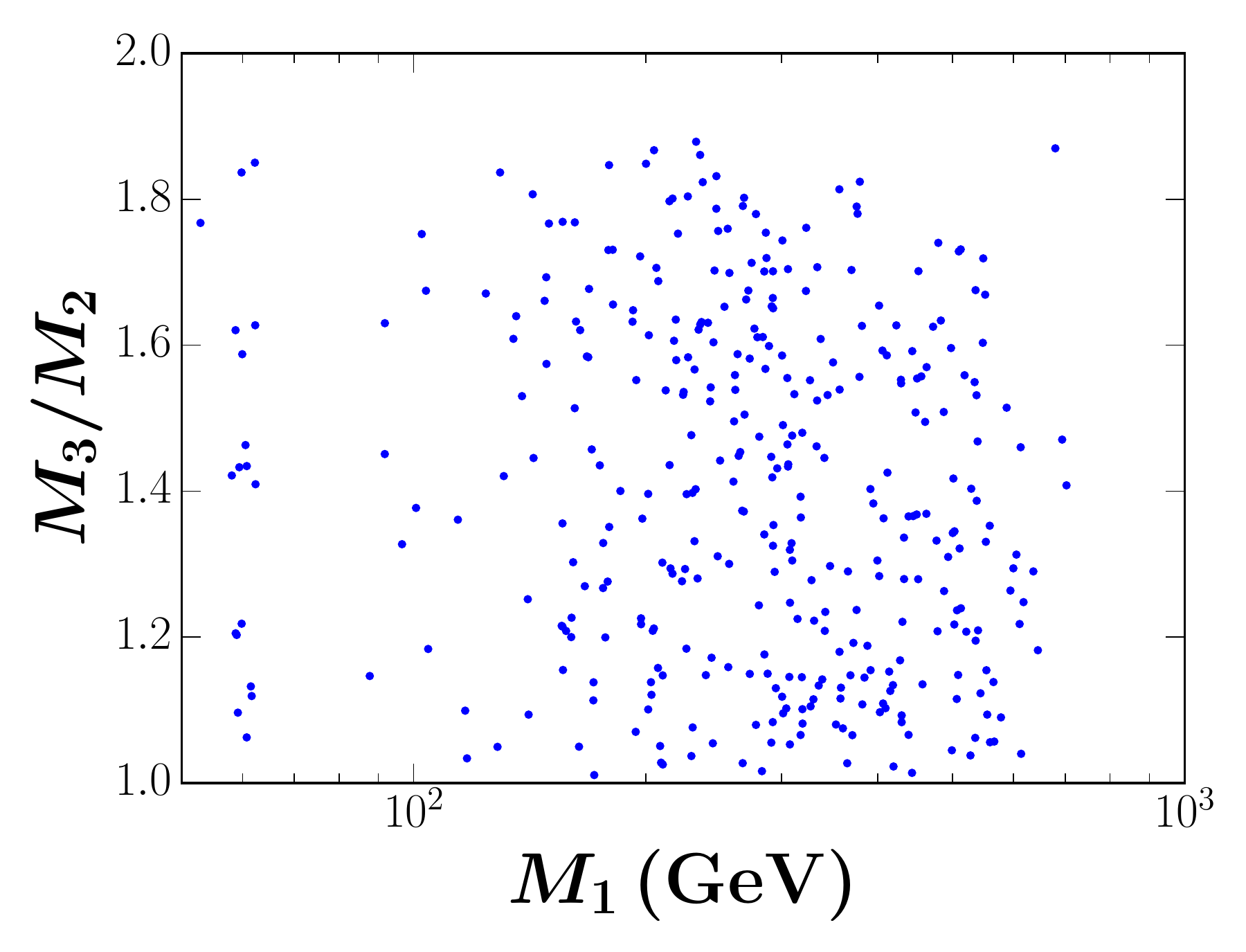}\\
\includegraphics[scale=0.4]{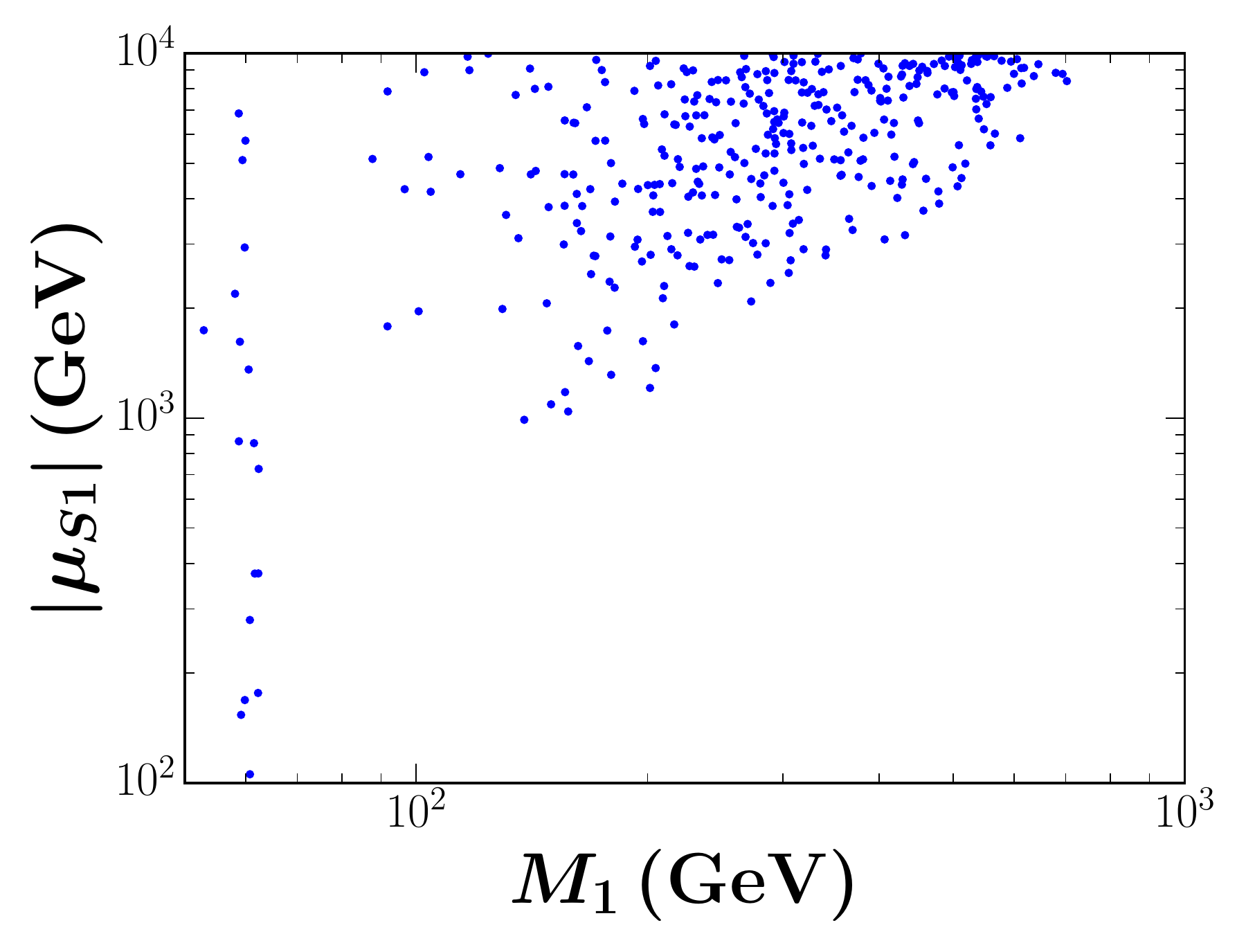}\hspace{1cm}
\includegraphics[scale=0.4]{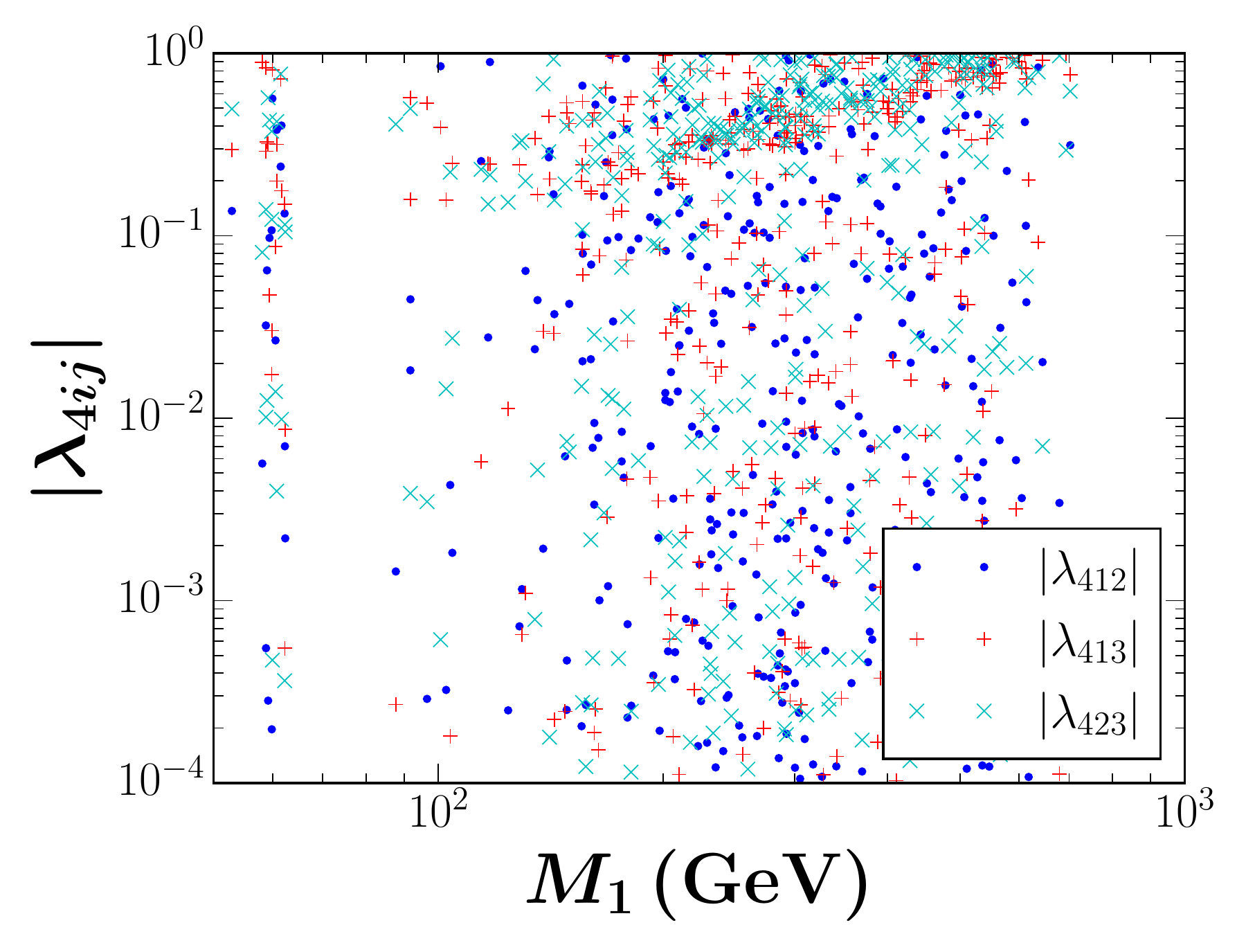}\\
\includegraphics[scale=0.4]{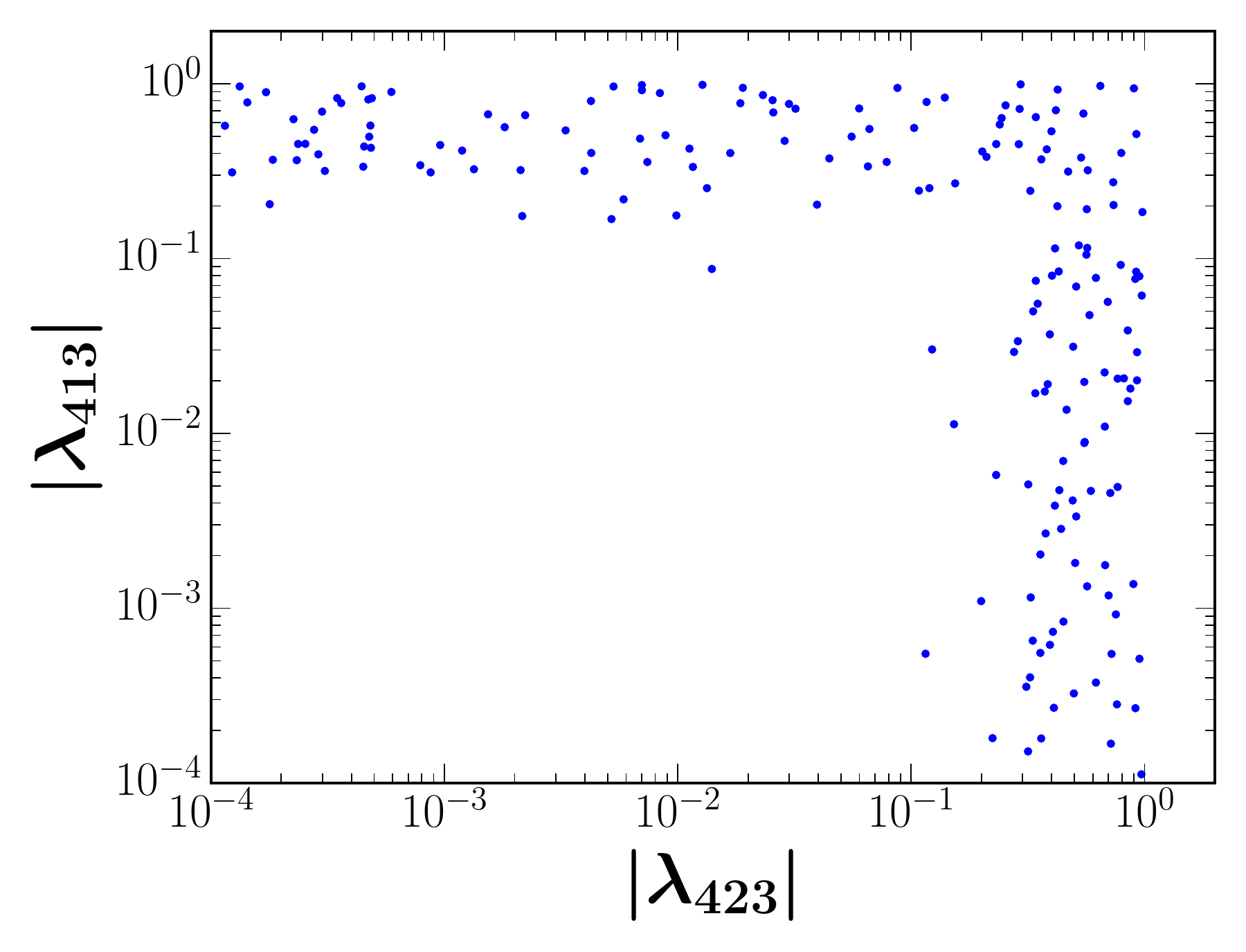}\hspace{1cm}
\includegraphics[scale=0.4]{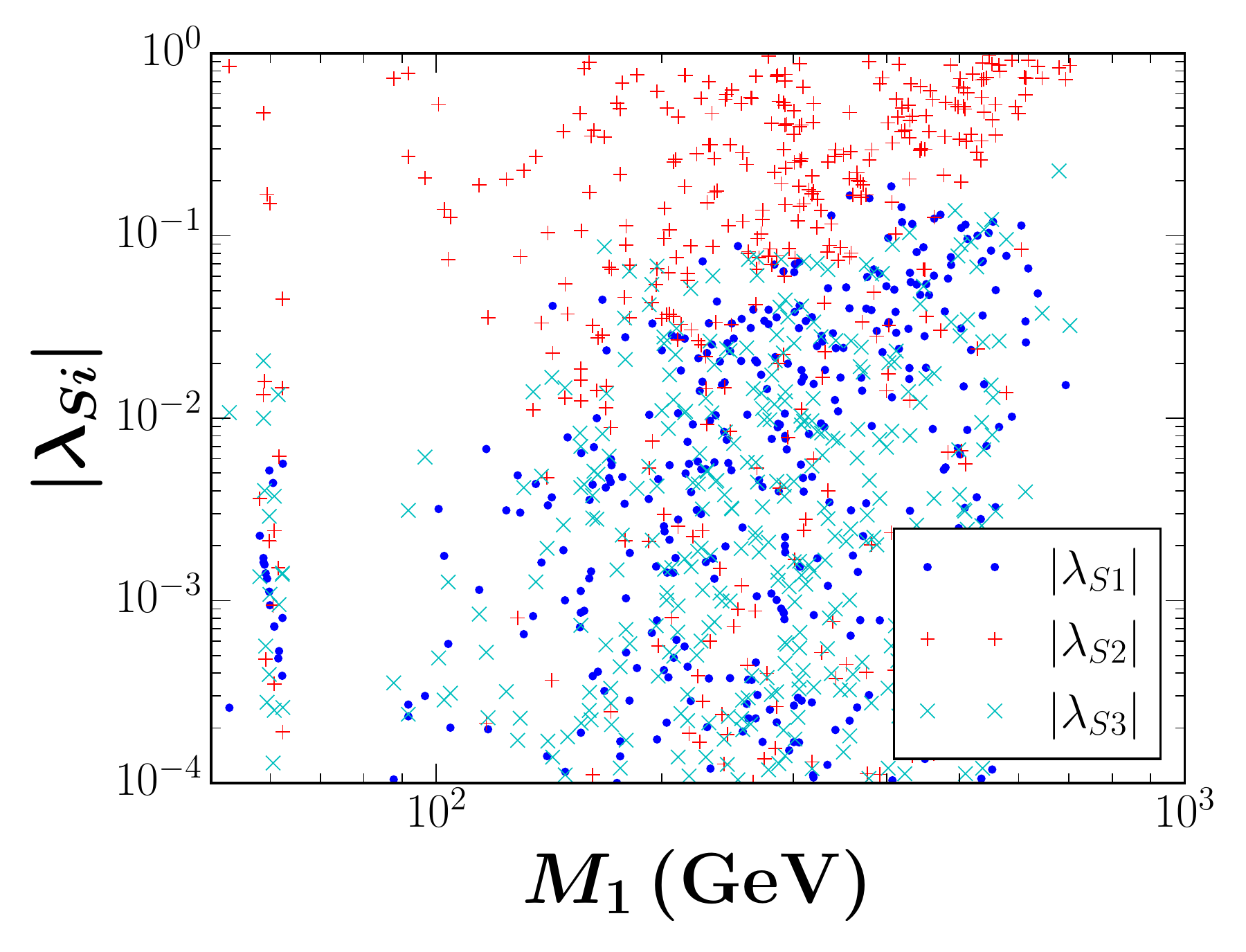}
\caption{The viable regions for $\mu_{S1}\neq0$ and $\lambda_{4ij}\neq0$ (scan C). All points shown in these plots are compatible with current data.}
\label{fig:scan-mus1-la4ij}
\end{figure}

For the third scan, we allow the quartic couplings $\lambda_{4ij}$ as well as the trilinear term $\mu_{S1}$ to randomly vary.  Since the $\mu_{S1}$ interaction does not affect $\phi_3$, its semiannihilation processes are absent and it is the conversions  induced by $\lambda_{4ij}$ 
($\phi_3\phi_3^\dagger\to \phi_i\phi_i^\dagger$) that play a key role in reducing $\Omega_3$ when $\lambda_{S3}$ is small. As we will see, this fact allows $\phi_3$, the heaviest dark matter particle, to give a significant contribution to the total dark matter density,  even the dominant one.  
The parameter $\mu_{S1}$ will induce semi-annihilation processes $\phi_1\phi_1 \to \phi_2 h$ that will reduce the abundance of $\phi_1$.

The viable parameter space for this case is illustrated in figure \ref{fig:scan-mus1-la4ij}. An important  novelty with respect to the two previous scans is that $\phi_1$ does not necessarily account for most of the dark matter --see top-left panel. In fact, either $\phi_3$ or $\phi_1$ could give the dominant contribution to the dark matter density, and their relic densities are often comparable. $\phi_2$, on the other hand, always gives a contribution below $5\%$ to the total relic density. The top-right panel shows that, as in the previous cases, the relic densities of the three dark matter components are  determined by conversions and semiannihilations. In particular for $\phi_1$, the coupling  $\mu_{S1}$ will induce semi-annihilation processes $\phi_1\phi_1 \to \phi_2 h$,  while annihilations do not play an important role. 
The next row shows that the ratios of masses $M_2/M_1$ and $M_3/M_2$ vary over a wide range, implying that  no degeneracies among the dark matter particles are required to satisfy the experimental bounds.  Regarding the couplings, $\mu_{S1}$ tends to increase with $M_1$, as expected --see bottom-left panel; the quartic couplings, meanwhile, can vary over several orders of magnitude, with $\lambda_{413}$ and $\lambda_{423}$ clustered toward higher values. These couplings are required to efficiently deplete the $\phi_3$ relic density via the conversions $\phi_3+\phi_3^\dagger\to \phi_{2,1}+\phi^\dagger_{2,1}$. Finally, from the bottom-center panel, we notice that, due to the direct detection constraints, $\lambda_{S1}$ and $\lambda_{S3}$ are both suppressed by an order of magnitude with respect to $\lambda_{S2}$. 

\begin{figure}
\centering
\includegraphics[scale=0.5]{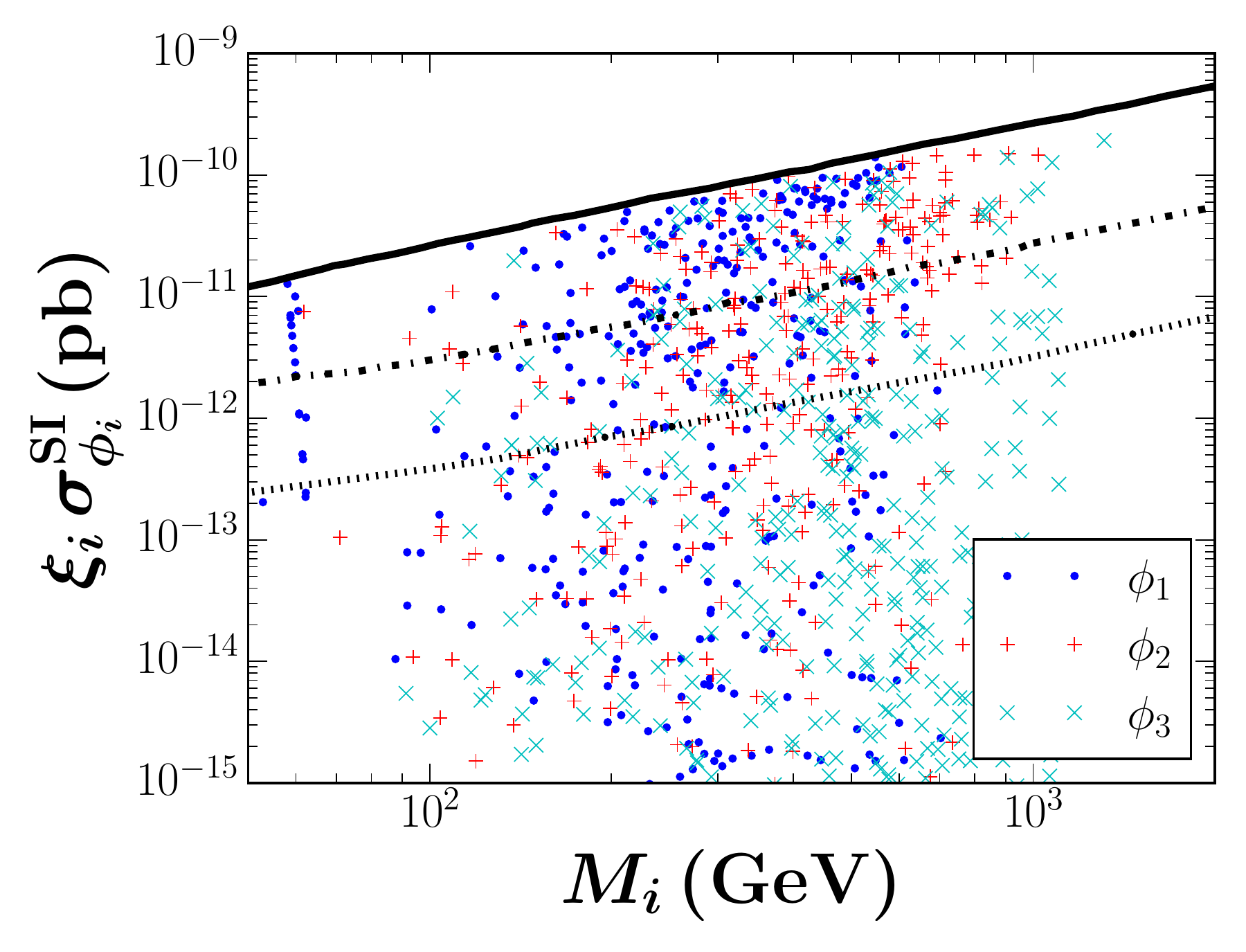}
\caption{Detection prospects in the scan with $\mu_{S1},\lambda_{4ij}\neq0$ (scan C).  The solid line is the current LZ limit whereas the dotted lines correspond to the expected sensitivities of LZ and DARWIN.} 
\label{fig:scan-mus1det}
\end{figure}

The detection prospects for this case are shown in figure \ref{fig:scan-mus1det}. Note that all three particles may give rise to observable signals in DARWIN. Since the dark matter masses are in general non-degenerate, one might be able to differentiate the three signals, excluding all vanilla models of dark matter and confirming the multi-component nature of the dark sector.

\subsection{$\lambda_{34}\neq0, \lambda_{4ij}\neq0, \mu_{S3}\neq0$}

\begin{figure}
\centering
\includegraphics[scale=0.4]{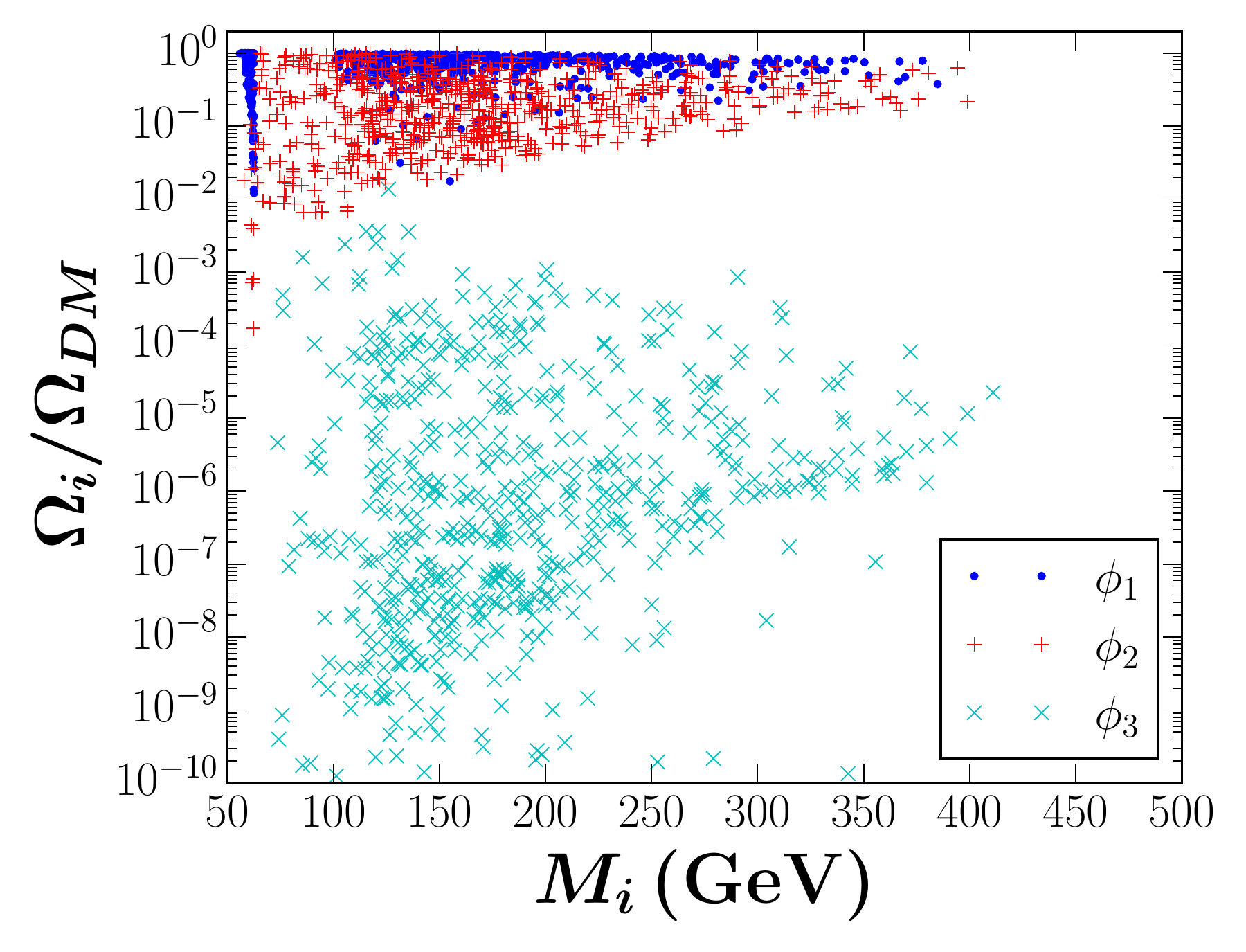}\hspace{1cm}
\includegraphics[scale=0.4]{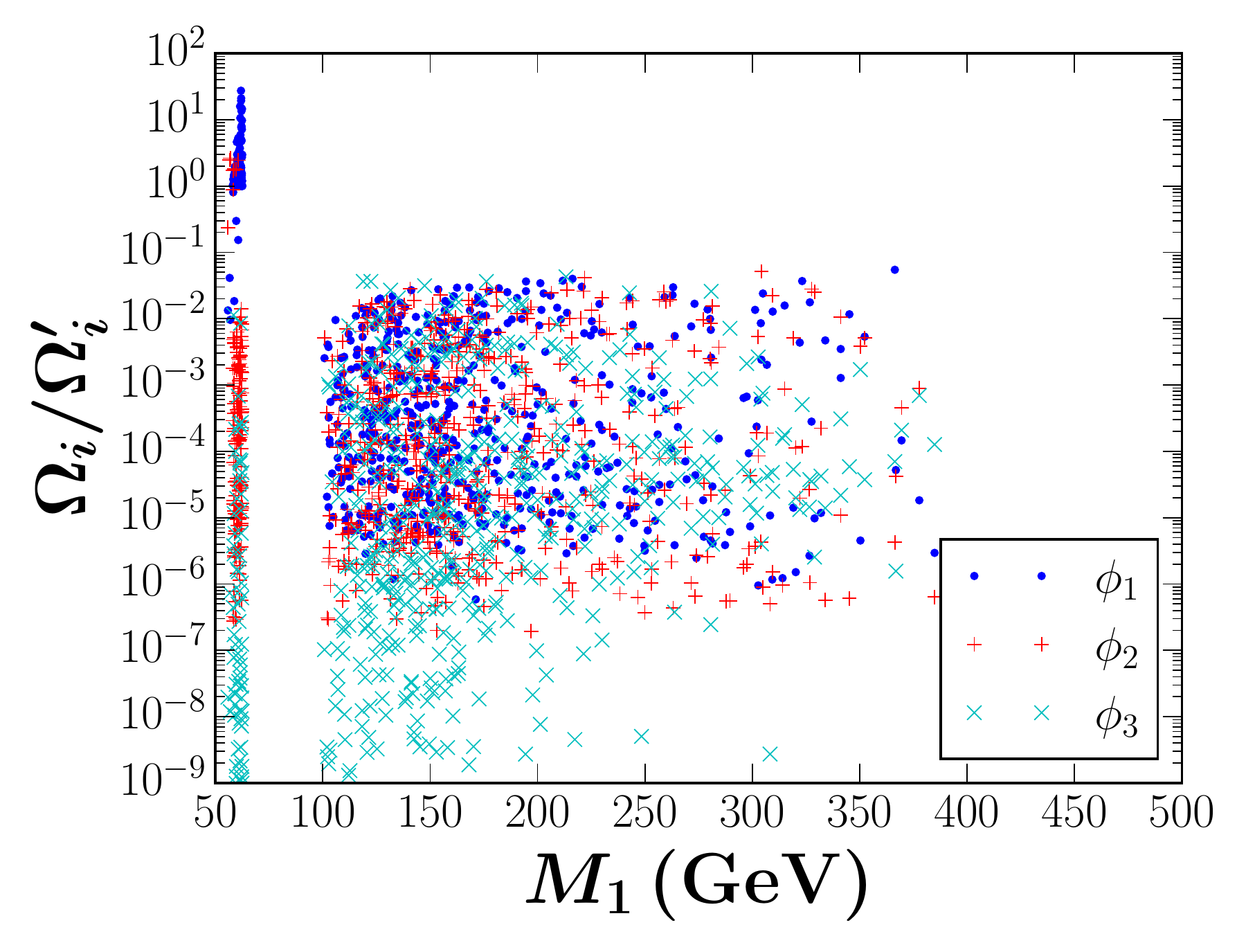}\\
\includegraphics[scale=0.4]{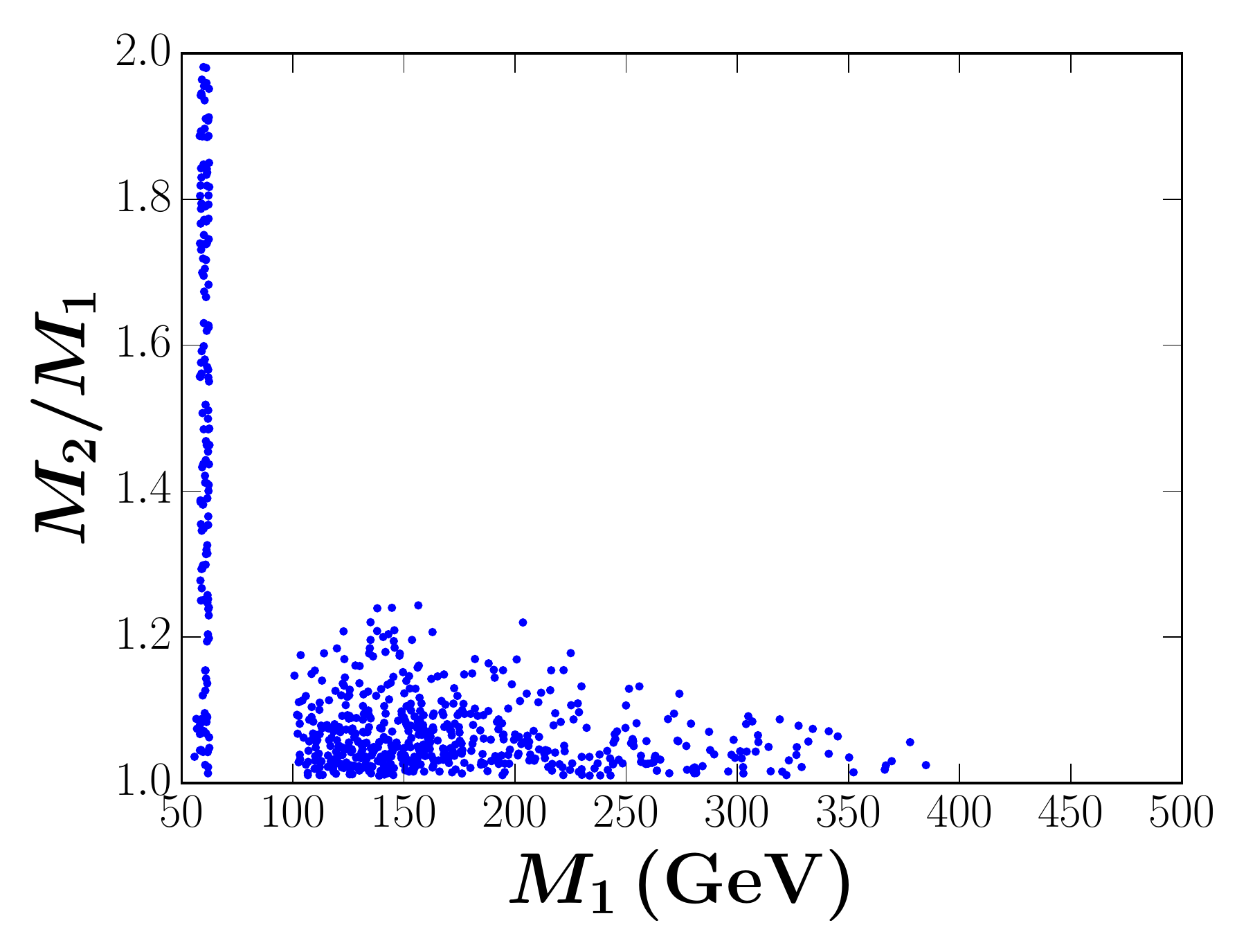}\hspace{1cm}
\includegraphics[scale=0.4]{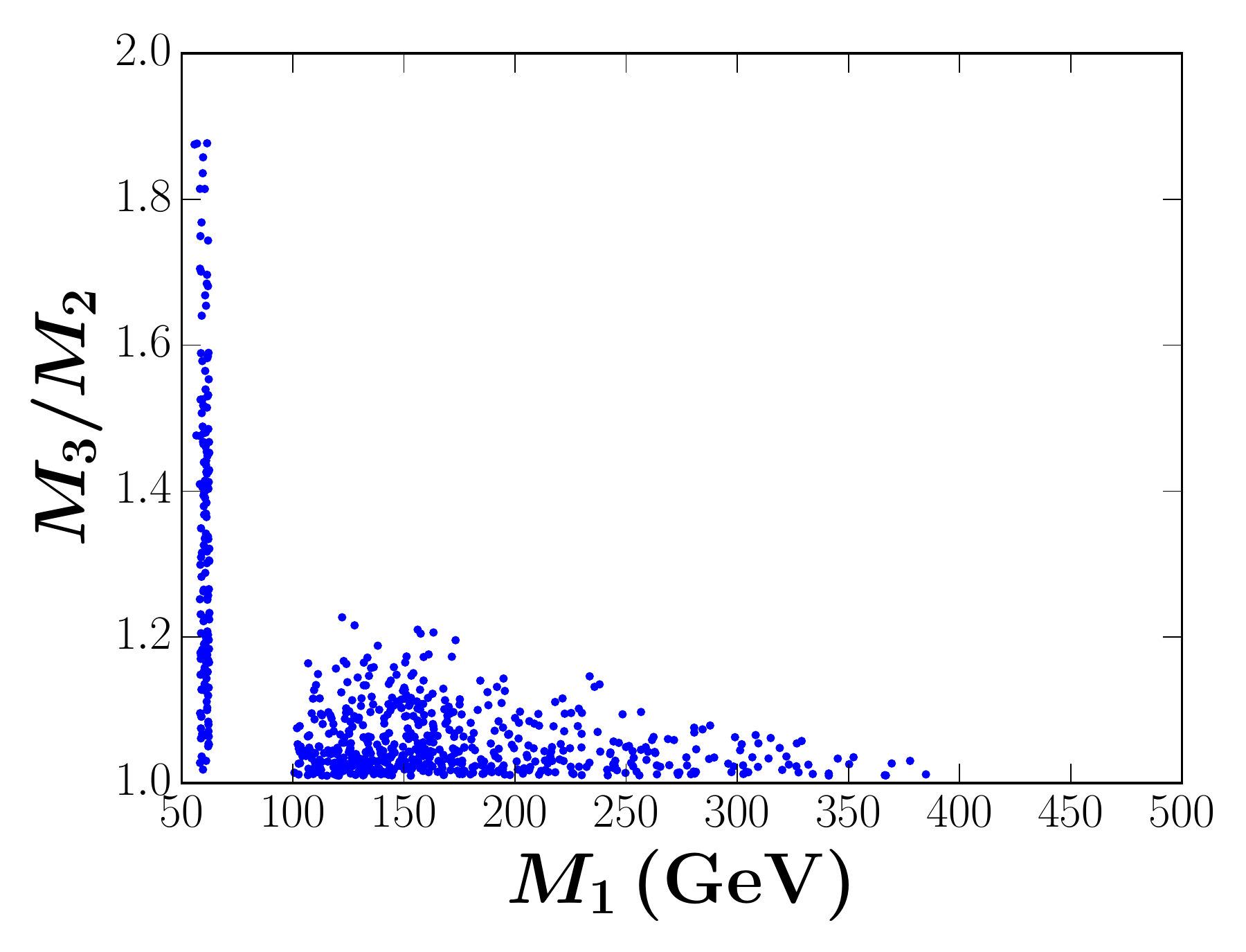}\\
\includegraphics[scale=0.4]{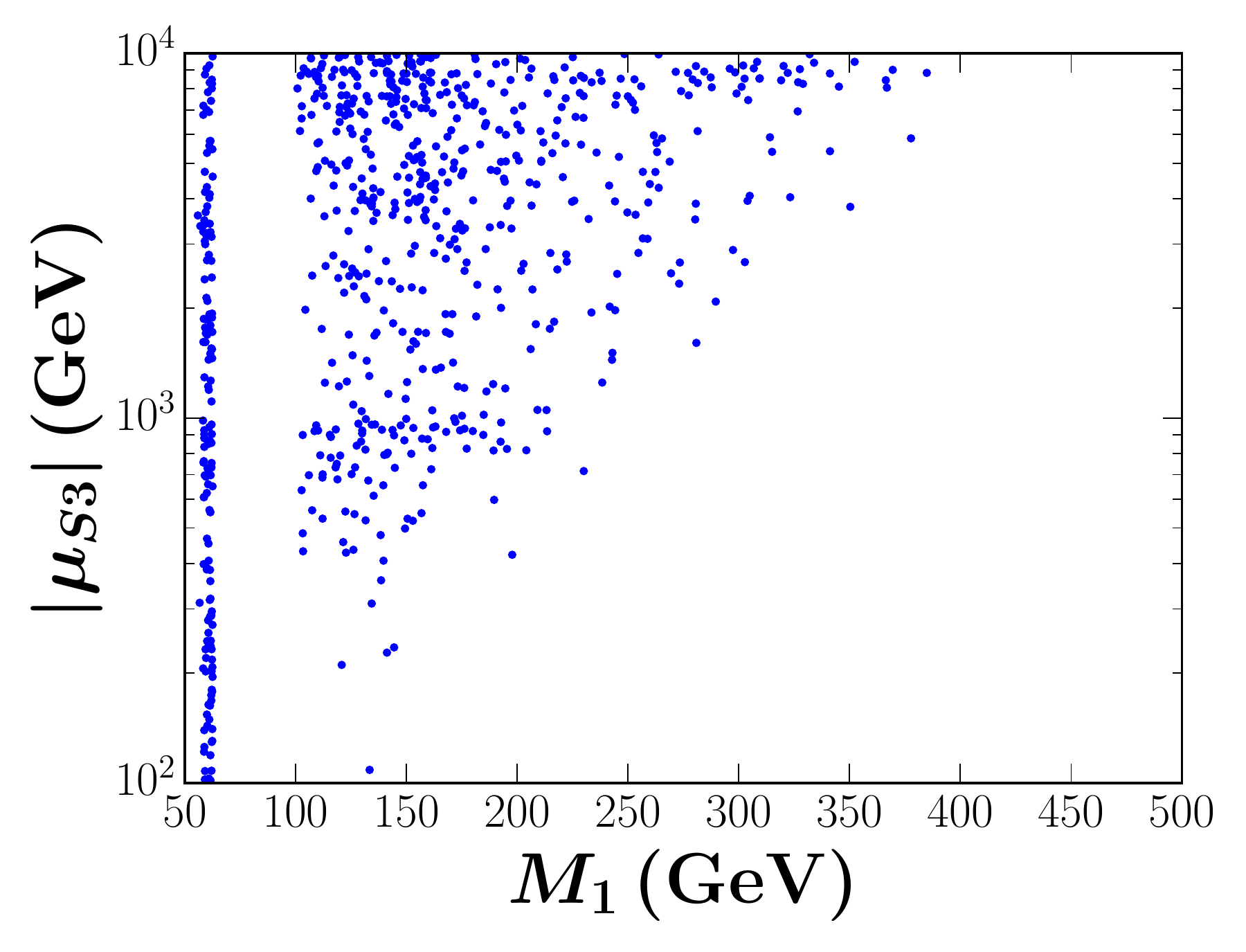}\hspace{1cm}
\includegraphics[scale=0.4]{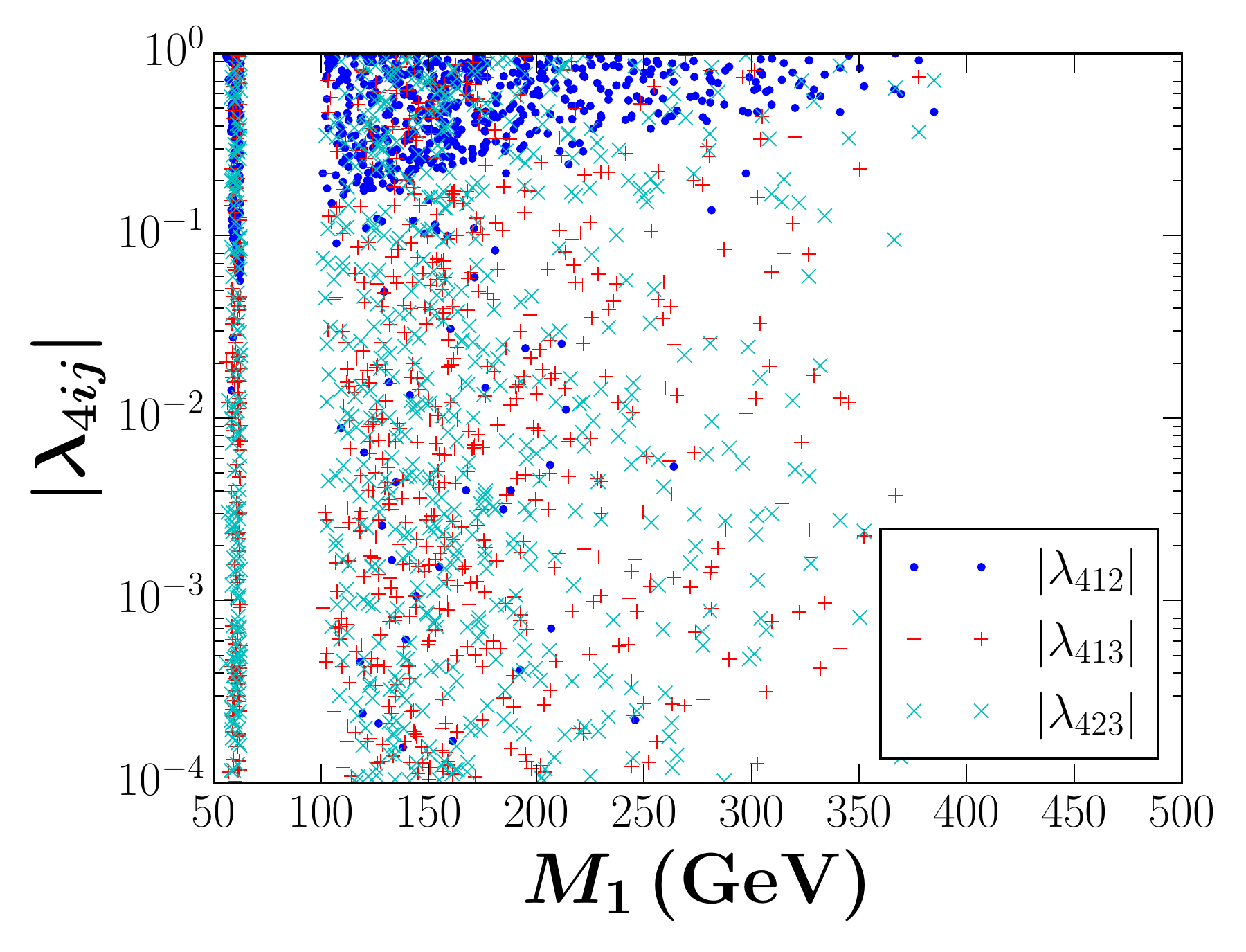}\\
\includegraphics[scale=0.4]{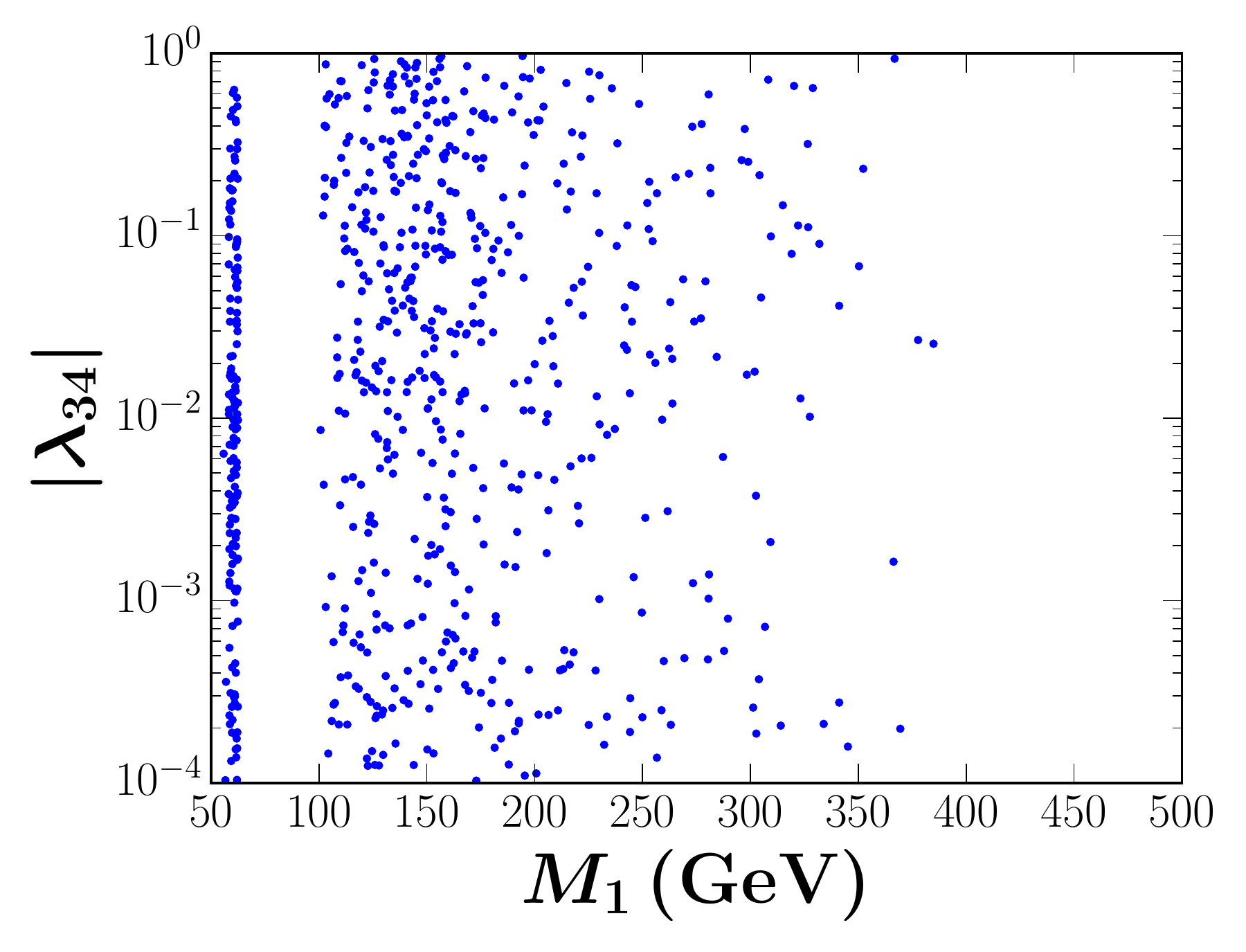}\hspace{1cm}
\includegraphics[scale=0.4]{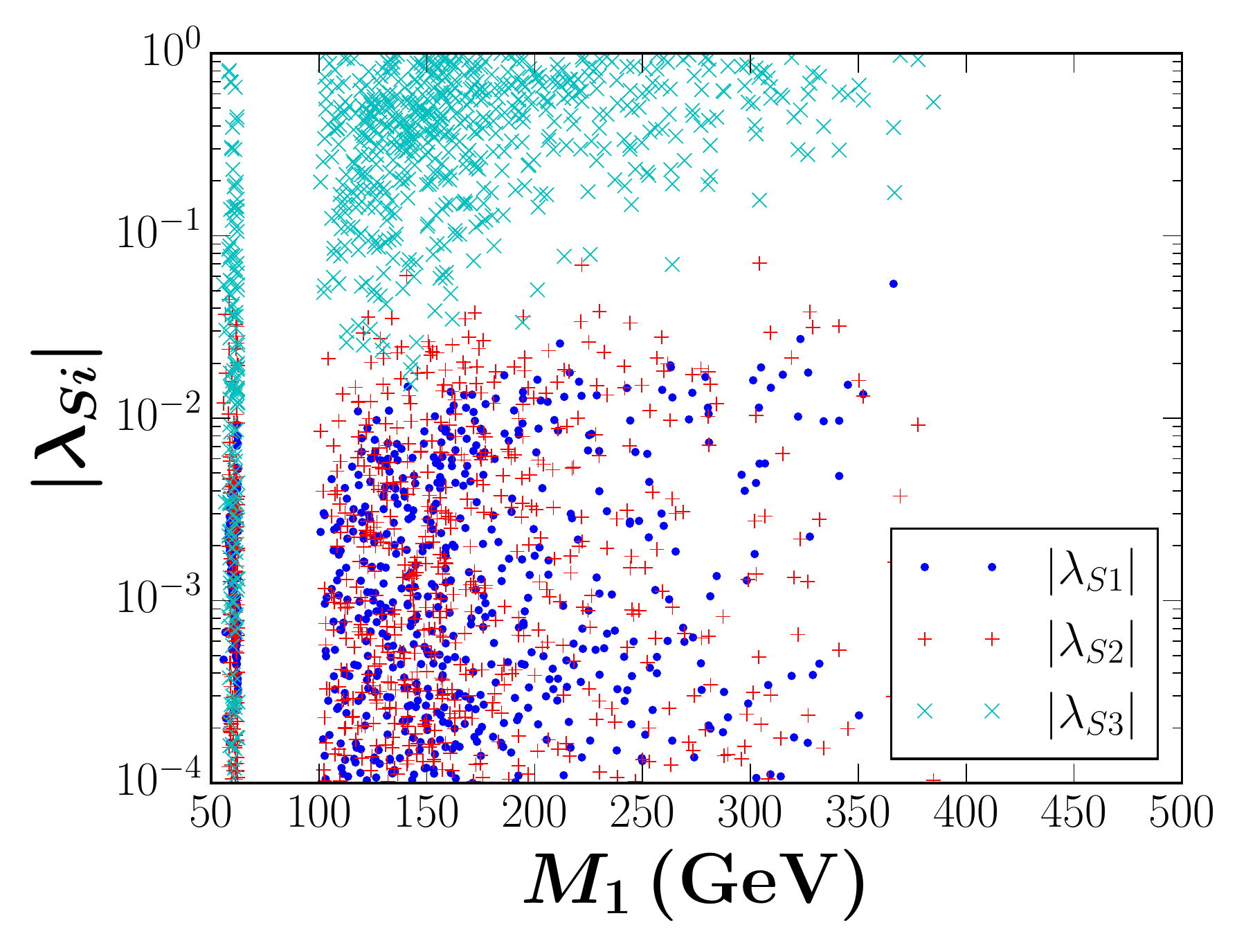}
\caption{The viable regions for $\mu_{S3}\neq0$, $\lambda_{Si}\neq0$,  $\lambda_{4ij}\neq0$ and $\lambda_{34}\neq0$ (scan D). All points shown in these plots are compatible with current data.}
\label{fig:scan-mus3-la34-la4ij}
\end{figure}

In the previous scans, the dominant contribution to the dark matter density was due to $\phi_1$ or $\phi_3$, so a natural question is whether $\phi_2$ might be responsible for the bulk of the dark matter in certain cases. With this scan we answer that question in the affirmative and we also incorporate into the discussion one of the $\lambda_{3i}$ couplings, which are specific to the $Z_7$ symmetry. Besides $\lambda_{34}$, the couplings that are different from zero are $\mu_{S3}$ and $\lambda_{4ij}$.

Figure \ref{fig:scan-mus3-la34-la4ij} displays the results for this scan, which, unlike  the previous ones, does not feature dark matter masses  above $400$ GeV or so. Thus, the dark matter masses are severely restricted in this case. Note that indeed the dark matter density is dominated by $\phi_1$ or $\phi_2$, but never by $\phi_3$, whose contribution is always below $1\%$ --see top-left panel. Once again dark matter conversions and semiannihilations play the dominant role in setting the relic density of the three components, we find that  $\Omega_i'<\Omega_i$ 
except when $M_1\approx m_h/2$ - see top-right panel. From the second row, we learn that the dark matter particles tend to be degenerate, with the ratios $M_2/M_1$ and $M_3/M_2$ rarely exceeding $1.2$ outside the Higgs-resonance, and decreasing rapidly with $M_1$.  The trilinear coupling $\mu_{S3}$ increases with $M_1$, reaching $10$ TeV (the maximum value allowed in the scan) at  $M_1\sim 400$ GeV, explaining why there are no viable models at higher masses.
This trilinear coupling enters the  semiannhilation processes  $\phi_1\phi_3 \rightarrow \phi_3^\dagger h$  and $\phi_3\phi_3 \rightarrow \phi_1^\dagger h$ which contribute to the relic density of $\phi_1$ and $\phi_3$ respectively. Conversion processes between the first and third sector, for example those involving the quartic coupling $\lambda_{34}$, also impact the relic density of $\phi_3$. All these processes lead to a very small value for $\Omega_3$, hence direct detection constraints can be evaded despite large values for $\lambda_{S3}$ - see bottom right panel.  The quartic coupling   $\lambda_{412}$ plays an important role as it induces the conversion process $\phi_2\phi_2^\dagger\to \phi_1\phi_1^\dagger$ that reduce the density of $\phi_2$ when the annihilation processes  $\phi_2\phi_2^\dagger\to hh,WW...$ are not efficient because of the small value of $\lambda_{S2}$.

\begin{figure}
\centering
\includegraphics[scale=0.5]{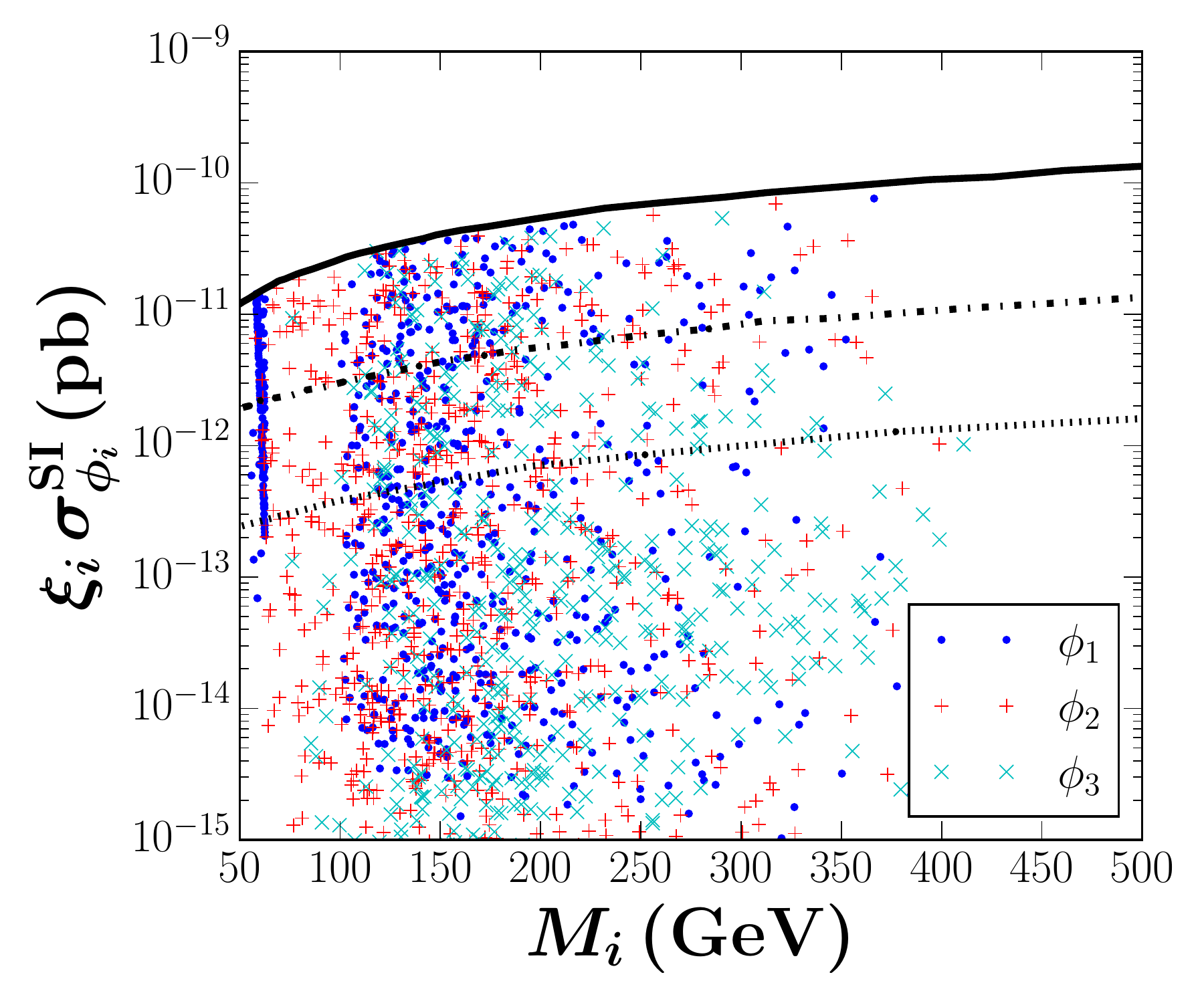}
\caption{Detection prospects in the scan with $\mu_{S3},\lambda_{34},\lambda_{4ij}\neq0$ (scan D). The solid line is the current LZ limit whereas the dotted lines correspond to the expected sensitivities of LZ and DARWIN.}
\label{fig:scan-mus3la34det}
\end{figure}

The direct detection prospects for this case are presented in figure \ref{fig:scan-mus3la34det}. There, we see that all three dark matter particles may give rise to observable signals in future direct detection experiments. Given the degeneracy of the masses, however, it is unlikely that their signals could be disentangled from one another, so as to establish the multi-component nature  of the dark matter. 

As a final remark, the results obtained for this scan hold even in the case when all the quartic couplings $\lambda_{3i}$ are allowed to vary in the same range. 

\section{Discussion}
\label{sec:disc}
\begin{figure}[t]
\centering
\includegraphics[scale=0.5]{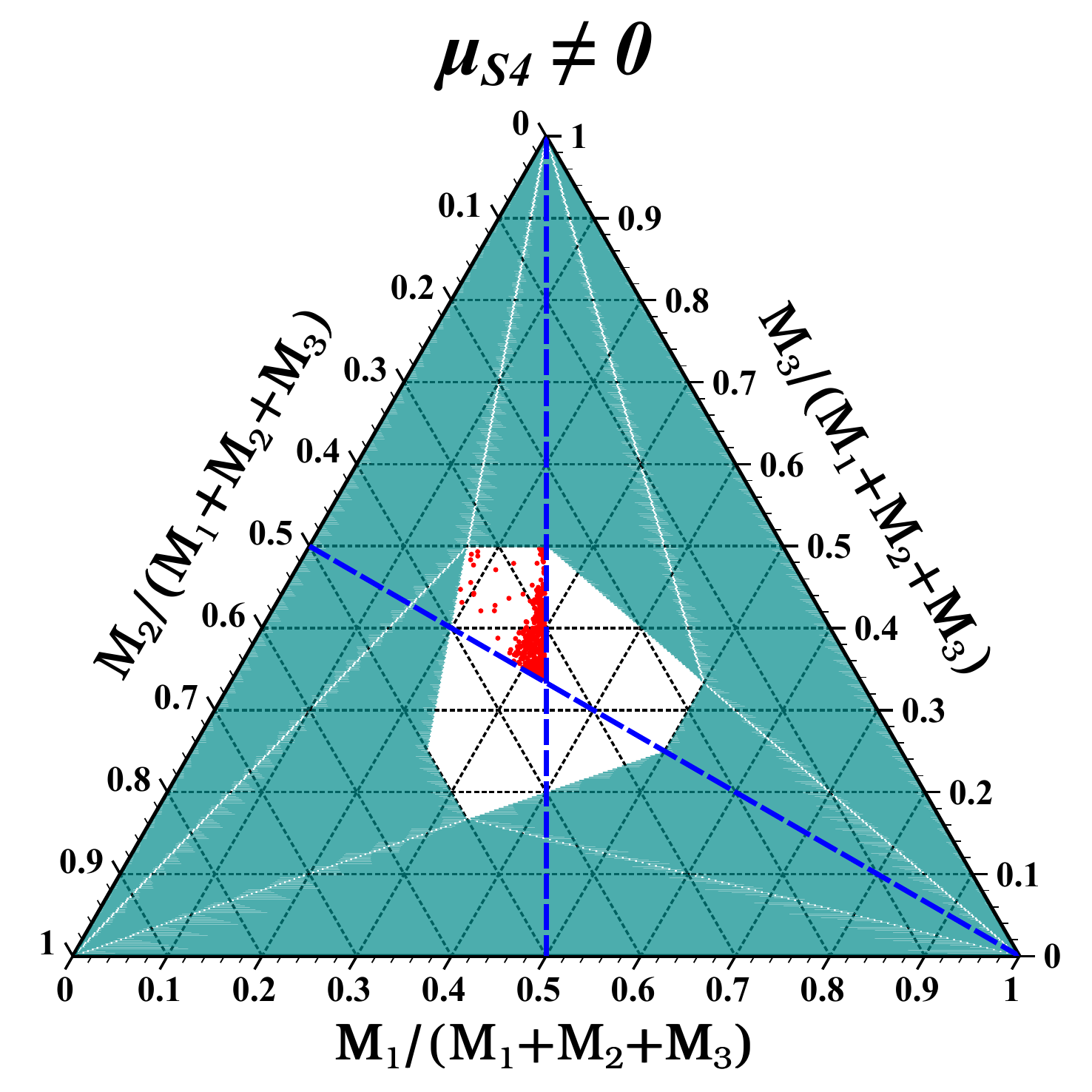}\hspace{1cm} \includegraphics[scale=0.5]{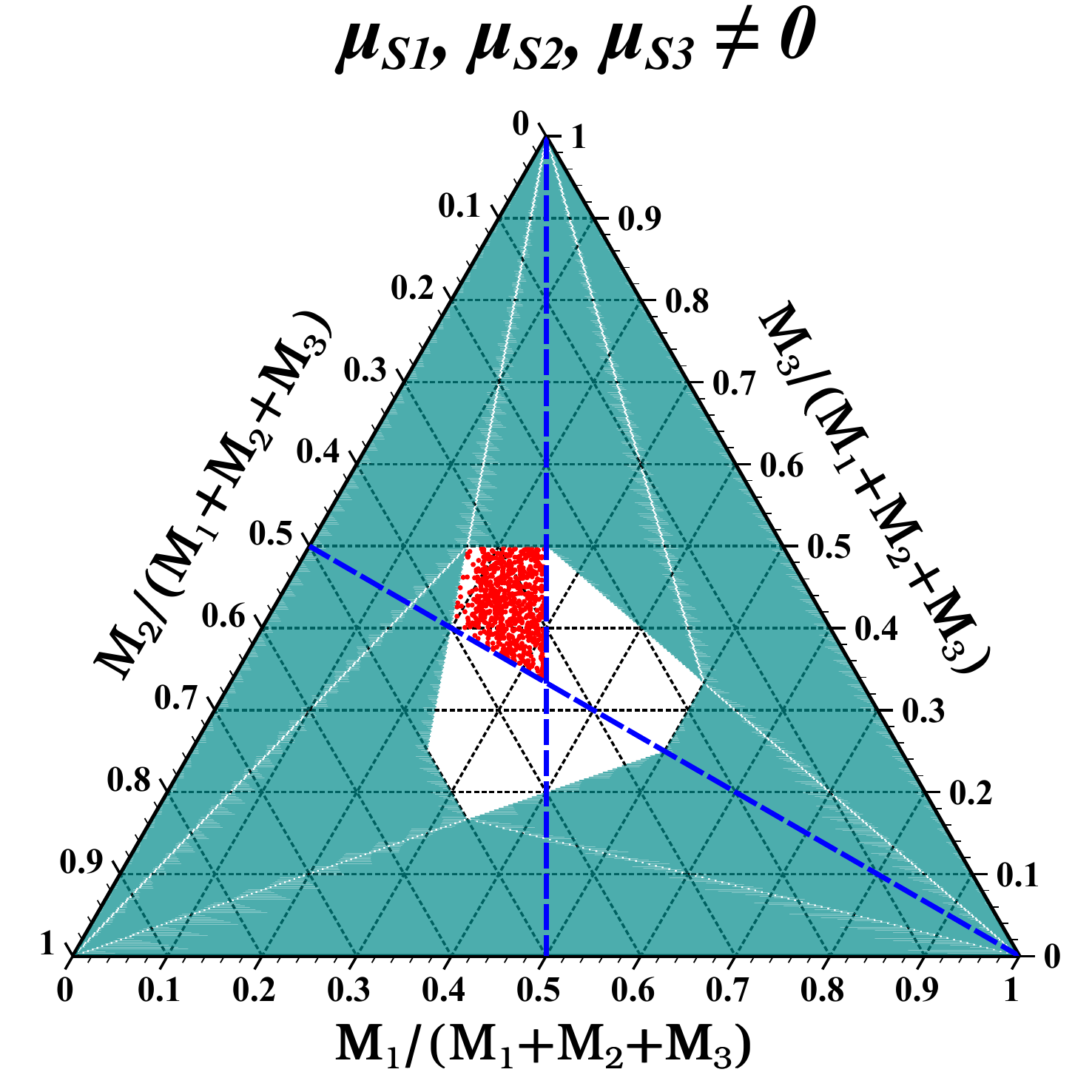}\\
\includegraphics[scale=0.5]{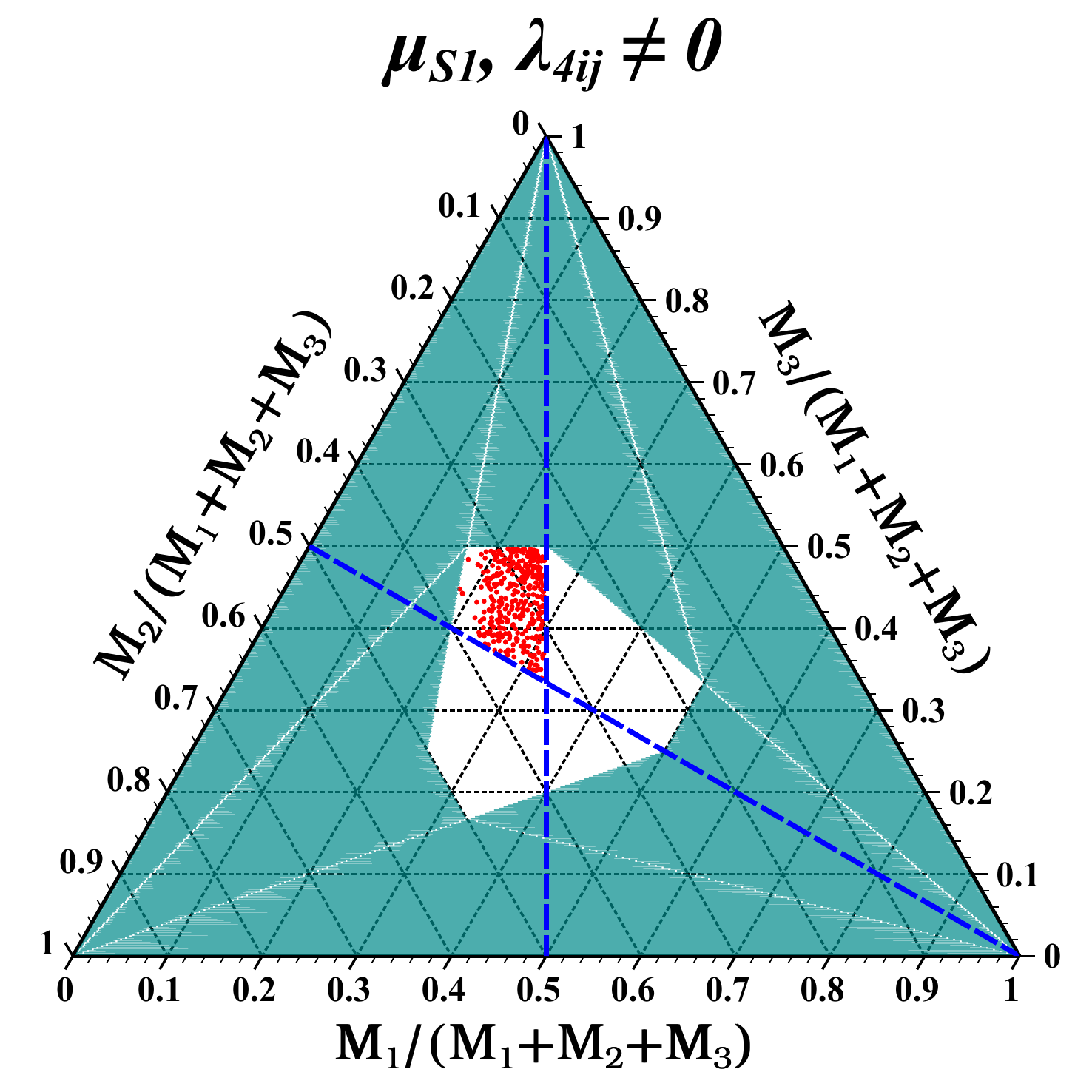}\hspace{1cm} \includegraphics[scale=0.5]{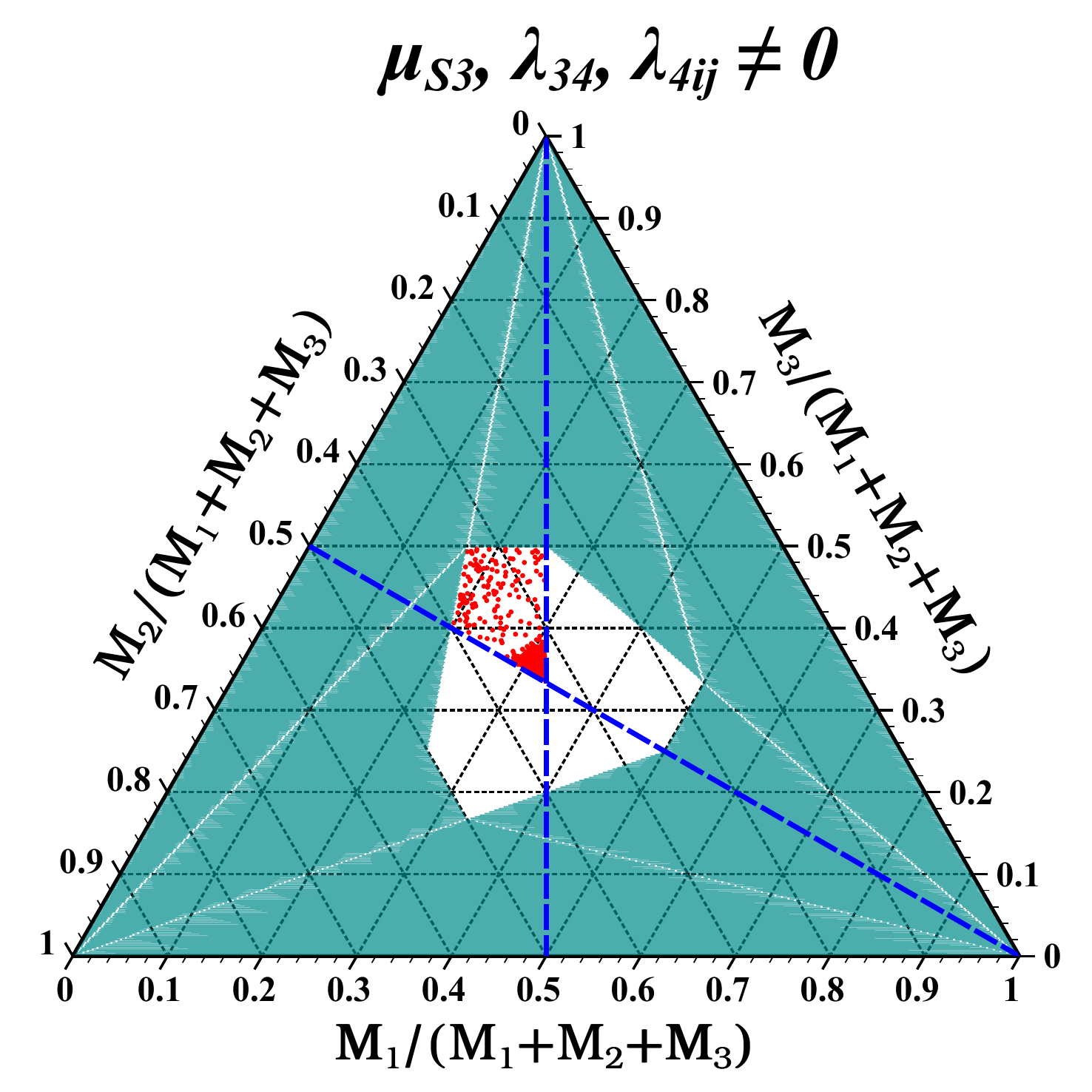}
\caption{Ternary plots showing the location of the viable models (red points) for each of the scans analyzed in the previous section. The choice $M_1<M_2<M_3$ restricts the possible location of the points to the top-left region limited by the two dashed lines. }
\label{fig:triangles}
\end{figure}

An illustrative way of comparing the results of the scans of the previous section is via the ternary plots shown in figure \ref{fig:triangles}. The three-component dark matter scenario considered in this work is realized in the central white region, with the choice $M_1<M_2<M_3$ further restricting the possible area to the top-left corner limited by the two dashed lines. All models, viable or not, necessarily lie in that corner. The question then is whether the viable models populate that entire corner, or just a portion of it.  From the figure we see that in scans $B$ (top-right) and $C$ (bottom-left) the viable points (in red)   cover most of that area, distributing themselves more or less uniformly over it. In contrast, in scans $A$ (top-left) and $D$ (bottom-right), the viable points tend to gather  toward the central point, leaving some unpopulated areas. These results are consistent with the fact, already mentioned, that the latter require some degeneracy among the dark matter masses, whereas the former do not.   Taking all the scans into consideration, we would conclude, then, that the viable points do span the full available region. 

All the scans studied in the previous section present a common characteristic, namely at least one trilinear interaction term involving $\phi_1$ is allowed, so as to facilitate the depletion of its relic density. The heavier dark matter particles can, instead, reduce their relic density via conversions  into the lighter ones, without a kinematic suppression. Consequently,  scans having only quartic interactions of the type $\lambda_{4ij}$ or $\lambda_{3i}$  do not lead to viable points unless a highly degenerate mass spectrum is invoked. That is the reason why we did not consider such possibilities.

We did carry out additional scans, which were not discussed in detail. For example, a  scan where  the only trilinear coupling varied is $\mu_2$ results in  a mass degeneracy between $\phi_1$ and $\phi_2$, independently of the rest of the couplings. In fact, since the $\mu_2$ interaction  gives rise to neither semiannihilations nor conversion processes reducing the number density of $\phi_1$, the conversion processes  mediated by the quartic couplings $\lambda_{412},\lambda_{31}$ and $\lambda_{36}$ (those affecting the number density of $\phi_1$ and $\phi_2$) must be efficient, implying a degeneracy. In another  scan, we considered the possibility of variations of the trilinear couplings $\mu_{S1}$ and $\mu_{S2}$.  In this case,  the relic densities can be significantly decreased,  leading to a viable and interesting scenario.  These results, however, are rather similar to those for the scan  $B$, where $\mu_{S3}$ is also varied. In summary, the scans we have considered in the previous section show the main phenomenological aspects of $Z_7$ the model, illustrating its versatility and capability as a viable scenario with three dark matter particles.

We have not examined in detail the possible signals in indirect detection experiments since it is unlikely they will  further restrict the viable parameter space.  Such  signals are expected to be similar to those found  in the $Z_N$ two-component scenarios \cite{Belanger:2020hyh,Yaguna:2021vhb}, where the semi-annihilation process $\phi_1+\phi_1\to \phi_2+h$ turns out to be the most relevant one, featuring a cross section that can reach values as high as $10^{-25}\,\mathrm{cm^3/s}$. If that value is extrapolated to the $Z_7$ model for  the processes $\phi_1+\phi_1\to \phi_2+h$ and $\phi_1+\phi_1\to \phi_3+h$, no additional constraints are found. 

Now we proceed to briefly compare our  results for the $Z_7$ model with those previously obtained for the $Z_5$ model of two-component dark matter \cite{Belanger:2020hyh}.  This model comprises the trilinear scalar interactions 
\begin{align}\label{eq:Z5lag}
 \mathcal{V}_{Z_5}&\supset\frac{1}{2}\tilde{\mu}_{S1}\phi^2_1\phi_2^{*} + \frac{1}{2}\tilde{\mu}_{S2}\phi_2^2\phi_1  + \text{H.c.}. 
 \end{align}
From this, two simple scenarios may arise, depending on the free parameter that is neglected.  When $\tilde{\mu}_{S1}\neq0$ (whose interaction term is analogous to $\mu_{S1}$ in the $Z_7$ model), it was found that the ratio $M_2/M_1$ varies over a wide range, and that the semi-annihilations entail that $\phi_1$ always gives the dominant contribution. These findings are in utter agreement, as concerns $\phi_1$ and $\phi_2$, with those obtained here in the scan  C ($\mu_{S1}\neq0$, $\lambda_{4ij}\neq0$). When $\tilde{\mu}_{S2}\neq0$, it was found instead that: $M_1$ becomes more restricted; the dark matter particles have to be degenerate; and the conversion process $\phi_1+\phi_1\to \phi_2+\phi_2$ induced by $\tilde{\mu}_{S2}$ plays a main role in decreasing $\Omega_1$. Since the analogous interaction in the $Z_7$ model turns to be $\mu_{S3}$ with the identification $\phi_2\to\phi_3$, the results  of the scan D ($\mu_{S3}\neq0$, $\lambda_{4ij}\neq0$, $\lambda_{34}\neq0$) match with those deduced from the $Z_5$ model. All these observations illustrate the compatibility of our results with those for $Z_N$ two-component scenarios.

Let us finally comment about the applicability of our results to other $Z_N$ scenarios with three dark matter particles.  The lowest $Z_N$ allowing simultaneous stability for three scalar fields is a $Z_6$ with the dark matter particles transforming  as $(\phi_1\sim w_6, \phi_2\sim w_6^2, \phi_3\sim w_6^3=-1)$, thus becoming a scenario with two complex fields and one real field. In this case the new trilinear and quartic interactions are given by~\cite{Yaguna:2019cvp}
\begin{align}
  \mathcal{V}_{Z_6} \supset&\,\frac{1}{2}\mu_{S1}\phi^2_1\phi_2^* + \frac{1}{3!}\mu_{32}\phi^3_2  + \mu_{S4}\phi_1\phi_2\phi_3 +\lambda'_{31}\phi_1^2\phi_2^2   + \lambda'_{32}\phi_2^2 \phi_3\phi_1^* + \lambda_{34}\phi _1 ^3\phi_3 + \text{H.c.}. 
\end{align}
It follows that this model shares the $\mu_{S1}$, $\mu_{S4}$ and $\lambda_{34}$ interaction terms with the $Z_7$ model. This means that the results of the scans A ($\mu_{S4}\neq0$) and C  ($\mu_{S1}\neq0$, $\lambda_{4ij}\neq0$) apply to the $Z_6$ model. Following the same line of thought, this conclusion extends to arbitrary $Z_N$ models \cite{Yaguna:2019cvp} with dark matter fields $[\phi_1,\phi_2,\phi_3]$, since the $\mu_{S1}$ and $\mu_{S4}$ trilinear interactions (along with the $\lambda_{4ij}$ quartic interactions) are always $Z_N$ invariant. Furthermore, since the $Z_8$, $Z_9$ and $Z_{10}$ models with a different set of dark matter fields can lead to scenarios having interaction terms of the type  $\mu_{S1}$, $\mu_{S4}$ or both, they are also viable scenarios in light of our results for the $Z_7$ model.     

\section{Conclusions}\label{sec:conc}
We proposed and investigated a novel and simple scenario that can explain the dark matter by extending the Standard Model with just two ingredients: a $Z_7$ discrete symmetry, and three complex scalar fields (the dark matter particles) that have different  charges under it. This three-component dark matter model intrinsically differs from the two-component scenarios considered so far due to the semiannihilation and conversion processes involving three different dark matter species, which had not been studied before.  After introducing the model and qualitatively  discussing its dark matter phenomenology,  we numerically studied its parameter space so as to determine the regions that are consistent with current data. To that end, a series of random scans were implemented, and the viable points were selected for further analysis. Our results clearly indicate that this model is indeed viable over a wide range of dark matter masses. Semiannihilations and dark matter conversions  turn out to both play important roles in determining the relic densities, and, depending on the parameters, any of the scalar fields could give the dominant contribution to the observed dark matter density.   We found that, for many viable points,   the three dark matter particles may be observed at future direct detection experiments, providing an avenue  to differentiate this scenario from the conventional one, and to test the multi-component nature of the dark matter. Finally, we argued that many of our results can be generalized to scenarios with a different $Z_N$ symmetry.  All in all, the $Z_7$ model of scalar dark matter was found to be a viable and testable scenario for physics beyond the Standard Model that can account for the dark matter of the Universe. 

\section*{Acknowledgments}
OZ and CY received funding from the Patrimonio Autónomo - Fondo Nacional de Financiamiento para la Ciencia, la Tecnología y la Innovación Francisco José de Caldas (MinCiencias - Colombia) grant 82315-2021-1080. 
OZ has received further funding from MinCiencias through the Grant 80740-492-2021, and has been partially supported by Sostenibilidad-UdeA and the UdeA/CODI Grant 2020-33177. CY would like to acknowledge support from the ICTP through the Associates Programe (2020-2025). The work of GB and AP was supported in part by the CNRS and RFBF, project number 20-52-15005.
The work of AP was carried out within the scientific program “Particle
Physics and Cosmology” of Russian National Center for Physics and Mathematics.

\bibliographystyle{apsrev4-1long}
\bibliography{references}

\end{document}